\documentclass[a4paper,fleqn,usenatbib]{mnras}
\usepackage{ae,aecompl}
\usepackage{graphicx}
\usepackage{amsmath}
\usepackage{amssymb}
\usepackage{longtable}
\usepackage{color}
\usepackage{booktabs}

\def\Swift{\textit{Swift}}
\def\keV{\rm{keV}}
\def\erg{\rm{erg}}
\def\s{\rm{s}}
\def\p{\rm{p}}

\title[Lower occurrence of X-ray flares in SN-GRBs]
{A lower occurrence rate of bright X-ray flares in SN-GRBs than
$z<1$ GRBs: evidence of energy partitions?}

\author[Mu et al.]
{\parbox{\textwidth}
{Hui-Jun Mu$^{1}$, Wei-Min Gu$^{1}$\thanks{E-mail: guwm@xmu.edu.cn},
Jirong Mao$^{2,3,4}$, Tong Liu$^{1}$, Shu-Jin Hou $^{5}$, \\
Da-Bin Lin$^{6}$, Junfeng Wang$^{1}$, Taotao Fang$^{1}$, En-Wei Liang$^{6}$}
\\
$^{1}$Department of Astronomy, Xiamen University, Xiamen, Fujian 361005, China\\
$^{2}$Yunnan Observatories, Chinese Academy of Sciences, 650011 Kunming, Yunnan Province, China\\
$^{3}$Center for Astronomical Mega-Science, Chinese
Academy of Sciences, 20A Datun Road, Chaoyang District, Beijing, 100012, China\\
$^{4}$Key Laboratory for the Structure and Evolution
of Celestial Objects, Chinese Academy of Sciences, 650011 Kunming, China\\
$^{5}$College of Physics and Electronic Engineering,
Nanyang Normal University, Nanyang, Henan 473061, China\\
$^{6}$Guangxi Key Laboratory for Relativistic Astrophysics,
Department of Physics, Guangxi University, Nanning 530004, China\\
}

\date{Accepted XXX. Received YYY; in original form ZZZ}

\pubyear{2018}

\begin{document}
\label{firstpage}
\pagerange{\pageref{firstpage}--\pageref{lastpage}}
\maketitle

\begin{abstract}
The occurrence rates of bright X-ray flares in $z<1$
gamma-ray bursts (GRBs) with or without observed supernovae (SNe)
association were compared. Our Sample~\textrm{I}: the $z<1$ long GRBs (LGRBs)
with SNe association (SN-GRBs) and with early \Swift/X-Ray Telescope (XRT)
observations, consists of 18 GRBs, among which only two
GRBs have bright X-ray flares.
Our Sample~\textrm{II}: for comparison, all the $z<1$ LGRBs
without observed SNe association and with early \Swift/XRT observations,
consists of 45 GRBs, among which 16 GRBs present bright X-ray flares.
Thus, the study indicates a lower occurrence rate of bright X-ray flares
in Sample~\textrm{I} (11.1\%) than in Sample~\textrm{II} (35.6\%).
In addition, if dim X-ray fluctuations are included as flares, then 16.7\% of
Sample~\textrm{I} and 55.6\% of Sample~\textrm{II} are found to have flares,
again showing the discrepancy between these two samples.
We examined the physical origin of these bright X-ray flares and found
that most of them are probably related to the central engine reactivity.
To understand the discrepancy, we propose that
such a lower occurrence rate of flares in the
SN-GRB sample may hint at an energy partition among the GRB, SNe, and X-ray
flares under a saturated energy budget of massive star explosion.
\end{abstract}


\section{Introduction} \label{sec:intro}

Gamma-ray bursts (GRBs) are known as the most luminous
electromagnetic explosion in the Universe
\citep[see][for reviews]{Piran2004,Meszaros2006,Kumar2015}.
For example, the total isotropic energy of GRB 160625B can reach
$\sim 10^{54}~\erg$ \citep{Wang2017}.
Observations show that long-duration, soft-spectrum GRBs (LGRB)
are associated with supernovae (SNe) Ib/c
\citep[see][for reviews]{van1999,Soderberg2006,Woosley2006,Della2007},
which are generally believed to originate from
the collapses of massive stars \citep{Woosley1993,MacFadyen1999,
Woosley2002,Heger2003,Zhang2004,Smartt2009,Woosley2012}.
The direct evidence of the GRBs with SNe association (SN-GRBs)
is revealed by \citet{Hjorth2012},
which classifies the SN-GRBs into five
grades as follows. Sample A: spectroscopic SNe;
Sample B: a clear light curve bump and some spectroscopic
evidence;
Sample C: a clear bump consistent with other SN-GRBs put at
the spectroscopic redshift;
Sample D: a bump, but the inferred SN properties are not fully
consistent with other SN-GRBs,
or the bump was not well sampled, or there is no spectroscopic
redshift for the GRB;
Sample E: a bump, either of low significance or inconsistent
with other SN-GRBs.
Following the spirit of the above classification,
\citet{Cano2017b} presented a quite comprehensive database compiled
of the observational and physical properties of the GRB prompt emission
and SN-GRBs, respectively, which consists of 46 SN-GRBs.

On the other hand, X-ray flares were observed by
\Swift/X-Ray Telescope (XRT) both in
long and short GRBs after the prompt gamma-ray emission
\citep{Burrows2005,Fan2005,Zhang2006,Nousek2006,Liang2006,
Falcone2006,OBrien2006}.
A few flares can occur even up to several days after the GRB trigger
\citep[e.g.,][]{Chincarini2007,Chincarini2010,Falcone2007}.
The physical origins of X-ray flares remain mysterious,
which may be related to the late-time activity of the central engine
\citep[e.g.,][]{Kumar2000,Perna2006,Dai2006,Lazzati2007,
Falcone2007,Maxham2009,Chincarini2010,Margutti2010},
or to the external shock \citep[e.g.,][]{Proga2006,Giannios2006,
Curran2008,Bernardini2011}. The steep decay was observed both in
the decay phase of flares and the prompt emission
\citep[e.g.,][]{Uhm2016,Jia2016,Mu2016b,Lin2017a,Lin2017b}.
Additionally, the high variabilities in the steep decay phase
may originate from the activities of the central engine
\citep[e.g.,][]{Proga2003,Lei2007,Liu2010,Zhang2016,Lin2016}.
A criterion was introduced to judge the physical origin of X-ray flares,
which is based on the relative variability flux and timescale
\citep[e.g.,][]{Ioka2005,Bernardini2011,Mu2016b}.
The external origin of the flares means that the flares are
related to afterglow variability.
On the contrary, the internal origin corresponds
to the late-time activity of the central engine.
Two well-known types of central engines are
the hyper-accreting stellar-mass black hole
\citep[e.g.,][]{Paczynski1991,Narayan1992,MacFadyen1999,Perna2006,
Luo2013,Liu2017}
and the millisecond magnetar
\citep[e.g.,][]{Usov1992,Duncan1992,Rees2000,Zhang2002,Dai2006,Metzger2015}.
For a SN-GRB with an X-ray flare from internal origin, the central engine
should account for three explosions, i.e., the supernovae, the prompt
gamma-ray emission, and the X-ray flare.

The main purpose of this work is to compare the occurrence rates
of X-ray flares in the $z<1$ GRBs with or without observed SNe
association, and investigate the physics if significant discrepancy exists
between these two rates. The remainder of this paper is organised as follows.
Sample selection is presented in Section~\ref{sec:sample}.
The main fitting procedure used for X-ray flare data is shown in Section~\ref{sec:fit}.
Occurrence rate and physical origin of bright X-ray flares
are investigated in Section~\ref{sec:CE}.
Discussion and conclusions are summarised in Section~\ref{sec:Conclusions}.

\section{Sample selection}\label{sec:sample}

A recent review paper, \citet{Cano2017b}, presented an up-to-date
progress report of the connection between LGRBs and
their accompanying SNe. Their sample consists of 46 SN-GRBs, which
are classified into five grades, as mentioned in Section~\ref{sec:intro}.
In this work, the bright X-ray flares in GRBs with
or without observed SNe association were studied.
Then, the 46 SN-GRBs in \Swift/XRT data were examined to
investigate the X-ray afterglow of these sources.
The following selection criteria were used to derive a sample of targets.
\begin{itemize}
  \item (1) The starting point is the sample of
  \Swift/XRT-detected GRBs. We picked only events observed
  by the \Swift/XRT. Thus, 19 GRBs without XRT follow-up
  observations can be removed (the removed sources can be found
  by referring to Table~\ref{T1}).
  \item (2) Since most flares occur in the early time
\citep[$t_{\p} \la 100~{\s}$,][]{Yi2016}, we chose the GRBs
  with early \Swift/XRT follow-up observations
  (trigger time $\la~300~\s $), and therefore eight GRBs were removed.
  \item (3) In addition, by taking into account the
  \Swift \ orbital constraint, an adequate X-ray afterglow observation
  in the early time ($100~\s \sim 1000~\s$)
  \footnote{\url{https://swift.gsfc.nasa.gov/archive/grb_table/}}
  was necessary. Thus, we removed three GRBs with the poor sampling
  in the early time.
\end{itemize}
Among the 46 SN-GRBs in \citet{Cano2017b}, there are 16 GRBs matching the
aforementioned three criteria. Moreover, two recent SN-GRBs, GRB 161219B/SN 2016jca\citep{Ashall2017,Cano2017a} and GRB 171205A/SN 2017iuk
\citep{Postigo2017,Prentice2017}, were added to our Sample~\textrm{I}.
Thus, Sample~\textrm{I} consists of 18 GRBs, among which three sources are
X-ray Flashes (XRFs) \footnote{\citet{Hjorth2012}
showed that the XRF population is likely associated with massive stellar death.
XRFs are included as ``low-luminosity" GRBs \citep{Cano2017b}.}.
There are a total of seven XRFs among all the 48 SN-GRBs, which are noted
in the first column of Table~\ref{T1}.
In addition, the enumerated list pertaining to our sample selection is 
reported in Table~\ref{T1}, where the related comments are shown in 
the fourth column.

\begin{table*}
\caption{The total 48 SN-GRBs are taken into account in this work,
among which the seven XRFs are noted in the first column.
The comments in the fourth column: 18 SN-GRBs in Sample~\textrm{I},
where the two bright X-ray flares and one dim
X-ray fluctuation are denoted as ``B" (bright) and ``D" (dim)
, respectively;
8 GRBs without early \Swift/XRT follow-up observations;
3 GRBs with the poor sampling in the early time;
19 GRBs without \Swift/XRT observations.
The definition of the grades of SN-GRBs are from \citet{Hjorth2012}
and \citet{Cano2017b}, which are shown in the fifth column:
Sample $\mathbf{A}$: spectroscopic SNe;
Sample $\mathbf{B}$: a clear light curve bump and some spectroscopic
evidence; Sample $\mathbf{C}$: a clear bump consistent with other
SN-GRBs putting at the spectroscopic redshift;
Sample $\mathbf{D}$: a bump, but the inferred SN properties are
not fully consistent with other SN-GRBs, or the bump was not
well sampled, or there is no spectroscopic redshift for the GRB;
Sample $\mathbf{E}$: a bump, either of low significance or inconsistent
with other SN-GRBs.
The SN-GRB references may refer to Table~1 of \citet{Kovacevic2014}
and Table~4 of \citet{Cano2017b}.}
\begin{tabular}{lcccc}
\toprule
$\rm{GRB}$   & $\rm{SNe}$ & z & Comments& Grade\\
\midrule
$050416\rm A^{\rm XRF}$	&	--	&	0.6528 	&	Sample~\textrm{I}	&	D	\\
$060218^{\rm XRF}$	&	2006aj	&	0.03342 	&	Sample~\textrm{I}	&	A	\\
060729	&	--	&	0.5428 	&	Sample~\textrm{I}	&	D	\\
060904B	&	--	&	0.7029 	&	Sample~\textrm{I}(B)	&	C	\\
070419A	&	--	&	0.9705 	&	Sample~\textrm{I}	&	D	\\
080319B	&	--	&	0.9371 	&	Sample~\textrm{I}	&	C	\\
081007	&	2008hw	&	0.5295 	&	Sample~\textrm{I}	&	B	\\
090618	&	--	&	0.5400 	&	Sample~\textrm{I}	&	C	\\
$100316\rm D^{\rm XRF}$	&	2010bh	&	0.0592 	&	Sample~\textrm{I}	&	A	\\
100418A	&	--	&	0.6239 	&	Sample~\textrm{I}	&	D/E	\\
101219B	&	2010ma	&	0.55185 	&	Sample~\textrm{I}	&	A/B	\\
111228A	&	--	&	0.7163 	&	Sample~\textrm{I}	&	E	\\
120422A	&	2012bz	&	0.2825 	&	Sample~\textrm{I}	&	A	\\
120729A	&	--	&	0.8000 	&	Sample~\textrm{I}	&	D/E	\\
130427A	&	2013cq	&	0.3399 	&	Sample~\textrm{I}	&	B	\\
130831A	&	2013fu	&	0.4790 	&	Sample~\textrm{I}(D)	&	A/B	\\
161219B	&	2016jca	&	0.14750 	&	Sample~\textrm{I}(B)	&	A	\\
171205A	&	2017iuk	&	0.0386 	&	Sample~\textrm{I}	&	A	\\
\midrule
$050824^{\rm XRF}$	&	--	&	0.8281 	&	no rapid follow-up	&	E	\\
091127	&	2009nz	&	0.49044 	&	no rapid follow-up	&	B	\\
101225A	&	--	&	0.8470 	&	no rapid follow-up	&	D	\\
111209A	&	--	&	0.67702 	&	no rapid follow-up	&	A/B	\\
111211A	&	--	&	0.4780 	&	no rapid follow-up	&	B/C	\\
130702A	&	2013dx	&	0.1450 	&	no rapid follow-up	&	A	\\
140606B	&	--	&	0.3840 	&	no rapid follow-up	&	A/B	\\
150518A	&	--	&	0.2560 	&	no rapid follow-up	&	C/D	\\
\midrule
050525A	&	2005nc	&	0.6060 	&	poor sampling	&	B	\\
120714B	&	2012eb	&	0.3984 	&	poor sampling	&	B	\\
150818A	&	--	&	0.2820 	&	poor sampling	&	B	\\
\midrule
970228	&	--	&	0.6950 	&	no XRT	&	C	\\
980326	&	--	&	--	&	no XRT	&	D	\\
980425	&	1998bw	&	0.00866 	&	no XRT	&	A	\\
990712	&	--	&	0.4330 	&	no XRT	&	C	\\
991208	&	--	&	0.7063 	&	no XRT	&	E	\\
000911	&	--	&	1.0585 	&	no XRT	&	E	\\
011121	&	2001ke	&	0.3620 	&	no XRT	&	B	\\
020305	&	--	&	--	&	no XRT	&	E	\\
020405	&	--	&	0.6899 	&	no XRT	&	C	\\
020410	&	--	&	--	&	no XRT	&	D	\\
$020903^{\rm XRF}$	&	--	&	0.2506 	&	no XRT	&	B	\\
021211	&	2002lt	&	1.0040 	&	no XRT	&	B	\\
030329A	&	2003dh	&	0.16867 	&	no XRT	&	A	\\
$030723^{\rm XRF}$	&	--	&	--	&	no XRT	&	D	\\
030725	&	--	&	--	&	no XRT	&	E	\\
$031203^{\rm XRF}$	&	2003lw	&	0.10536 	&	no XRT	&	A	\\
040924	&	--	&	0.8580 	&	no XRT	&	C	\\
041006	&	--	&	0.7160 	&	no XRT	&	C	\\
130215A	&	2013ez	&	0.5970 	&	no XRT	&	B	\\

\bottomrule
\end{tabular}
\label{T1}
\end{table*}

\begin{figure*}
\begin{tabular}{cccc}
\includegraphics[angle=0,scale=0.18,trim=10 0 10 0,clip]{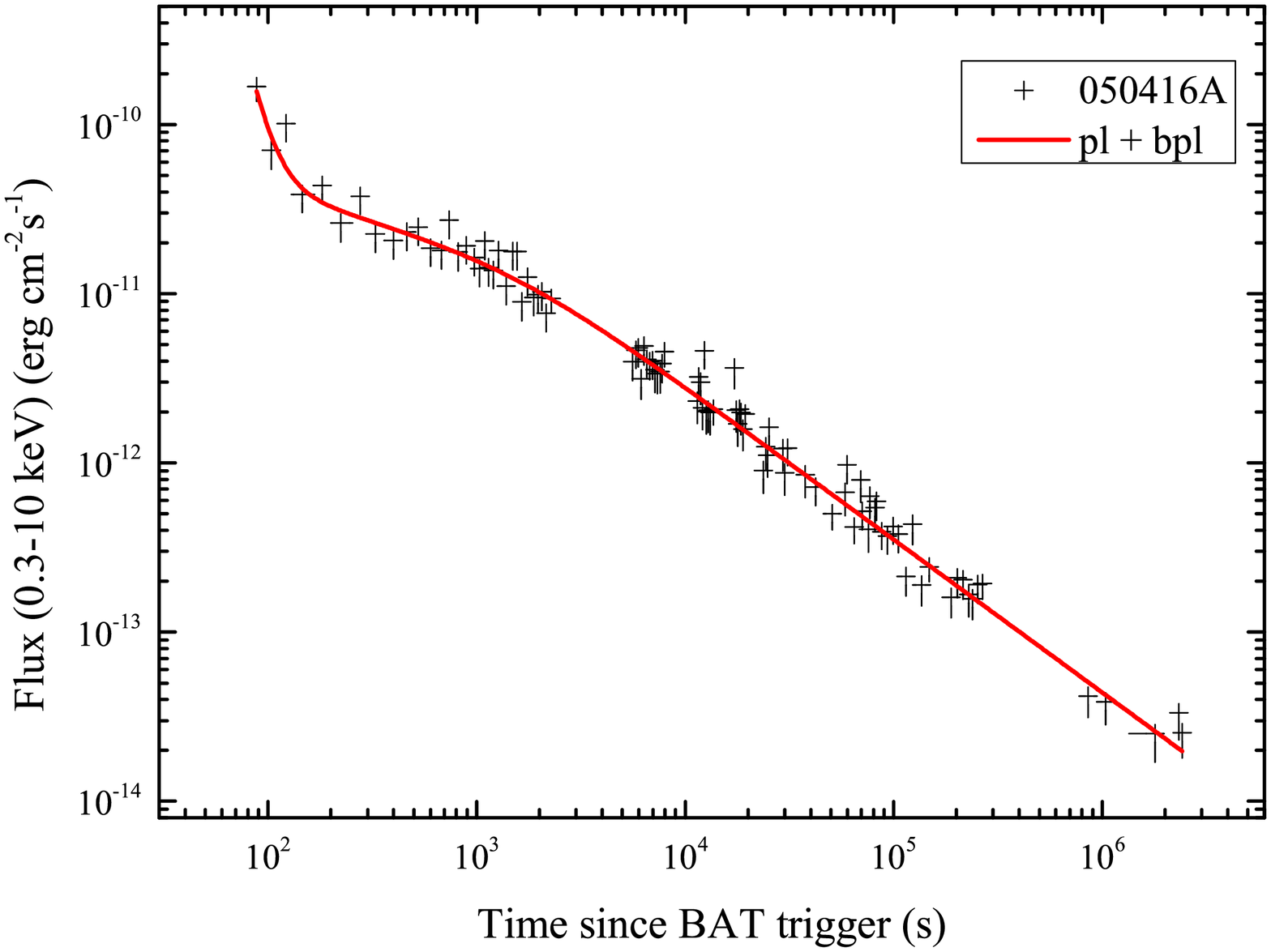}&
\includegraphics[angle=0,scale=0.18,trim=10 0 10 0,clip]{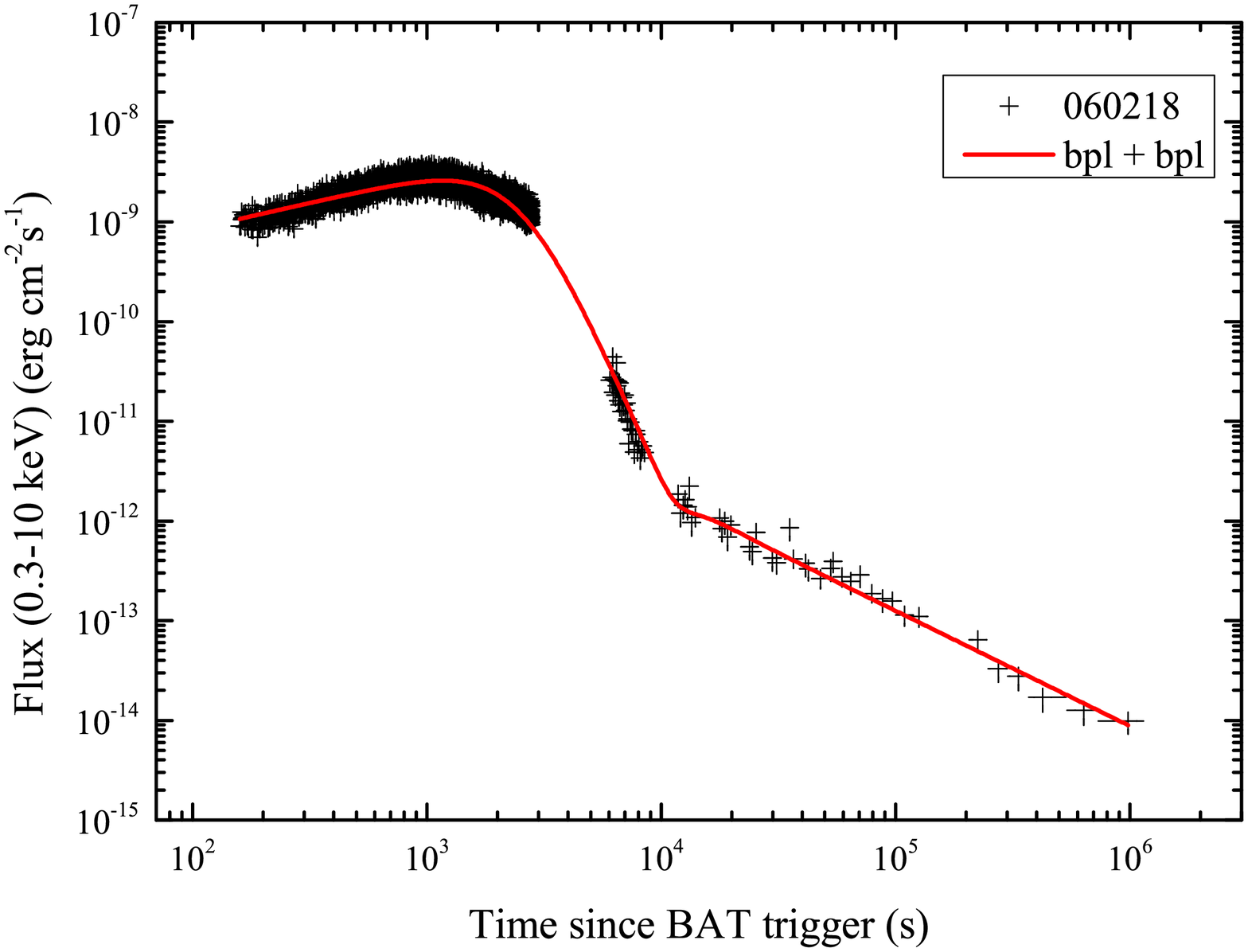} &
\includegraphics[angle=0,scale=0.18,trim=10 0 10 0,clip]{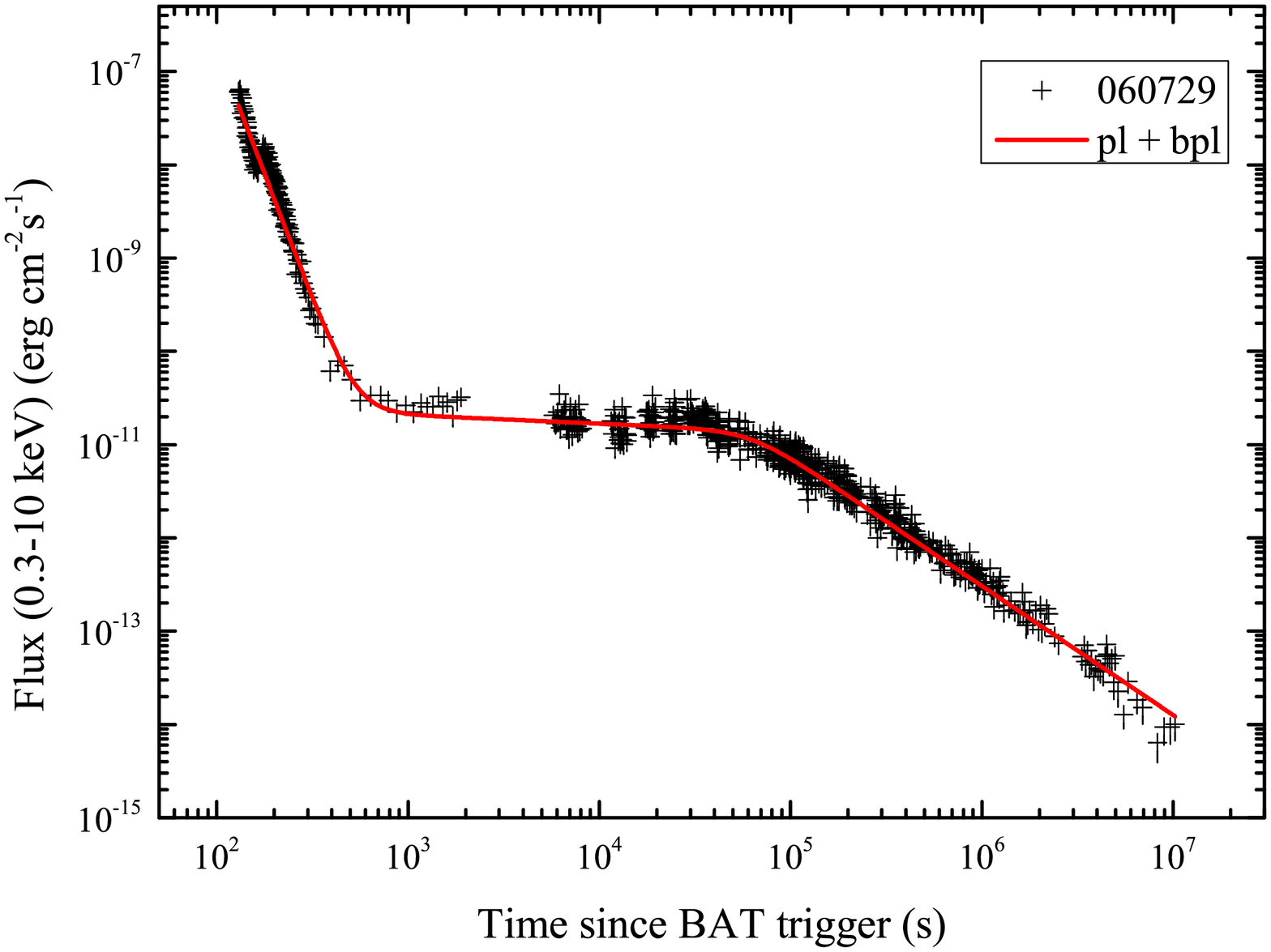}  \\
\includegraphics[angle=0,scale=0.18,trim=10 0 10 0,clip]{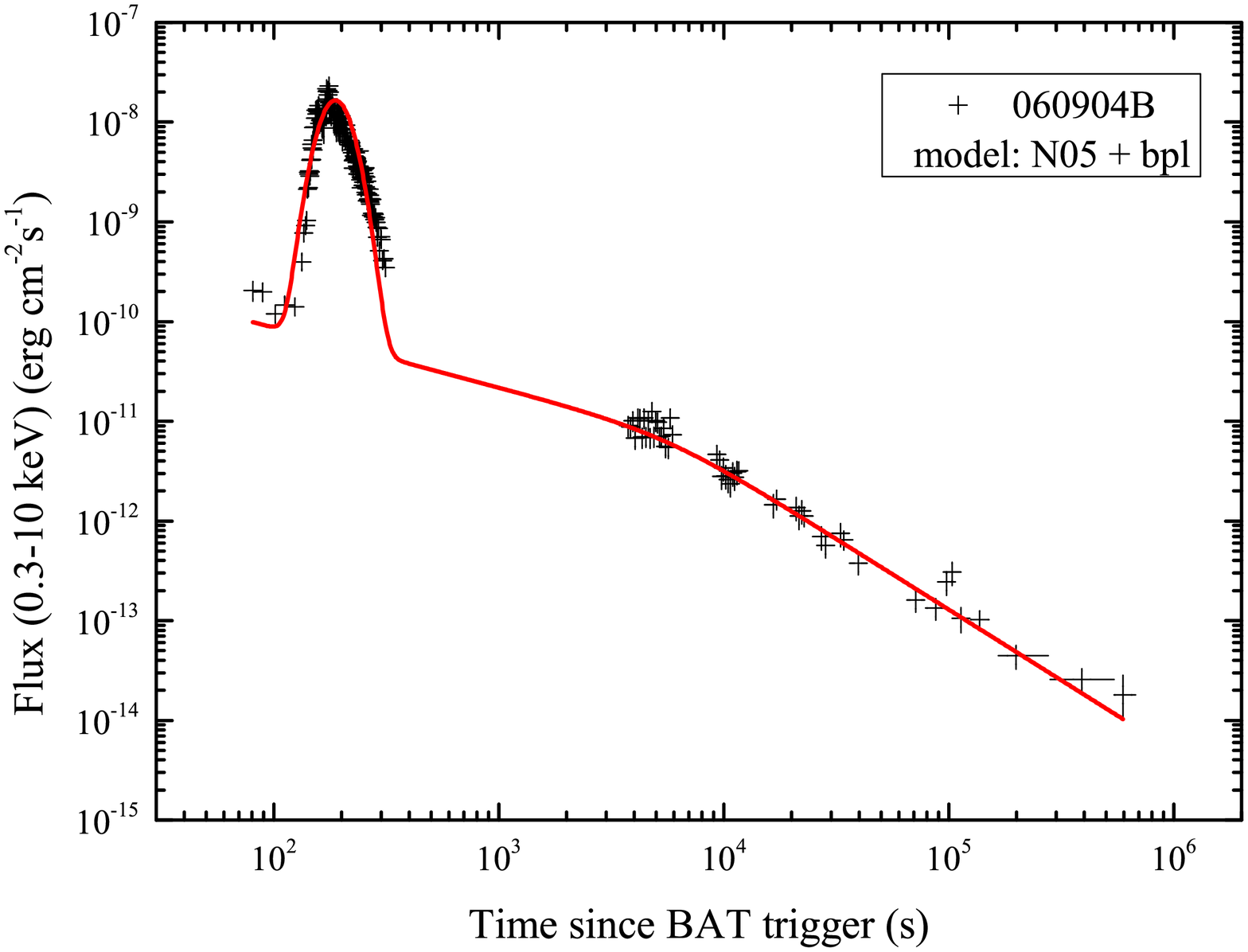} &
\includegraphics[angle=0,scale=0.18,trim=10 0 10 0,clip]{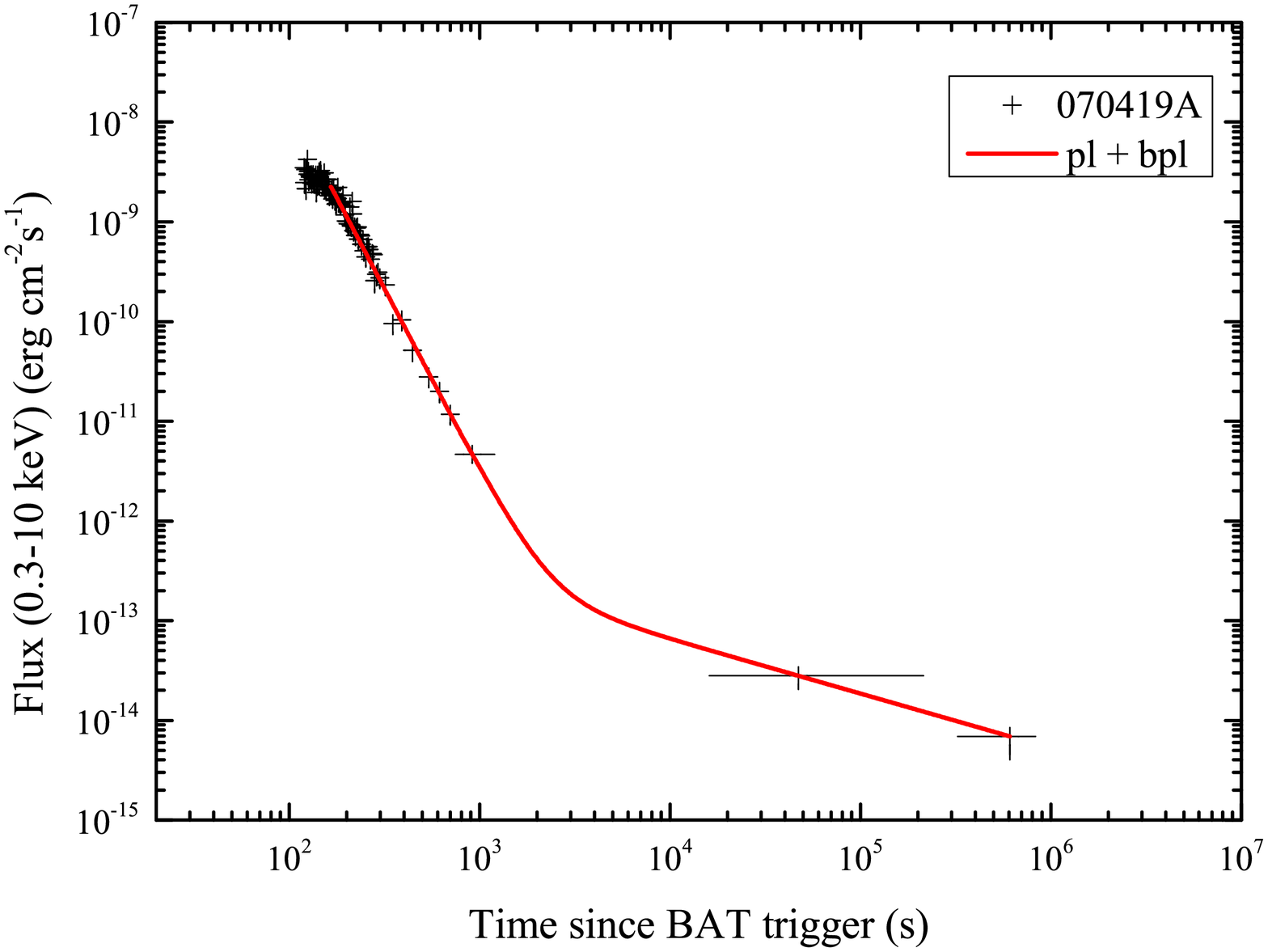} &
\includegraphics[angle=0,scale=0.18,trim=10 0 10 0,clip]{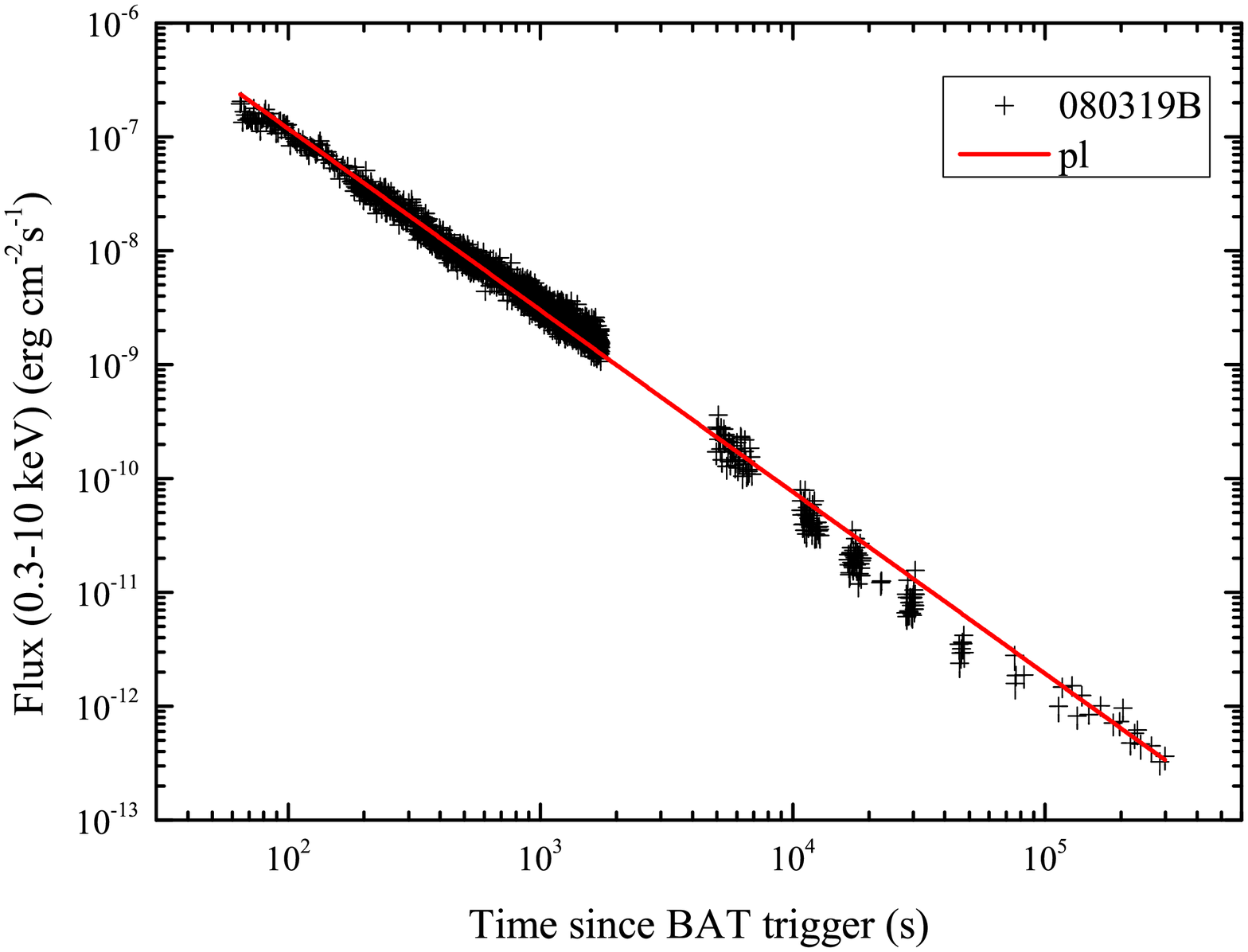} \\
\includegraphics[angle=0,scale=0.18,trim=10 0 10 0,clip]{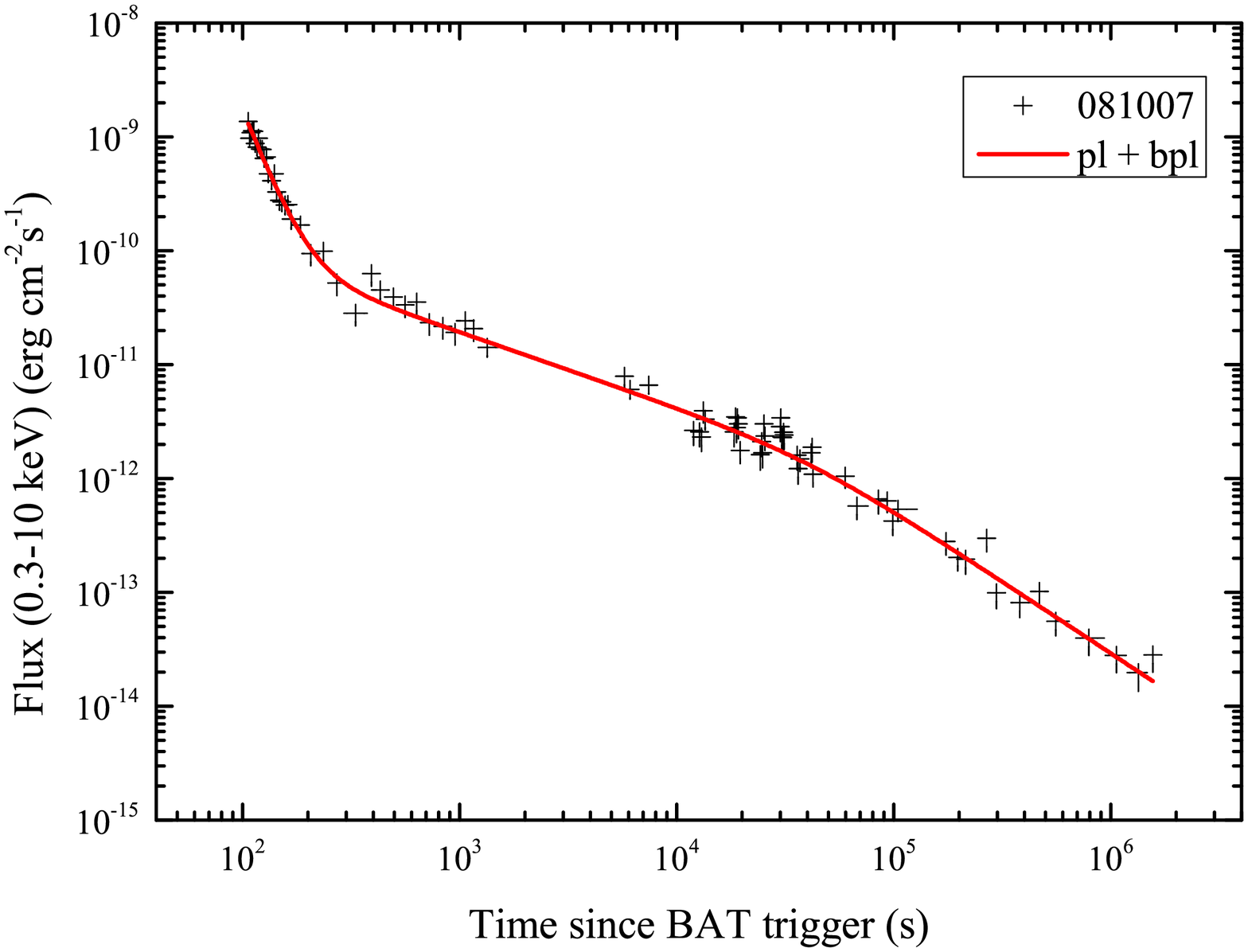}  &
\includegraphics[angle=0,scale=0.18,trim=10 0 10 0,clip]{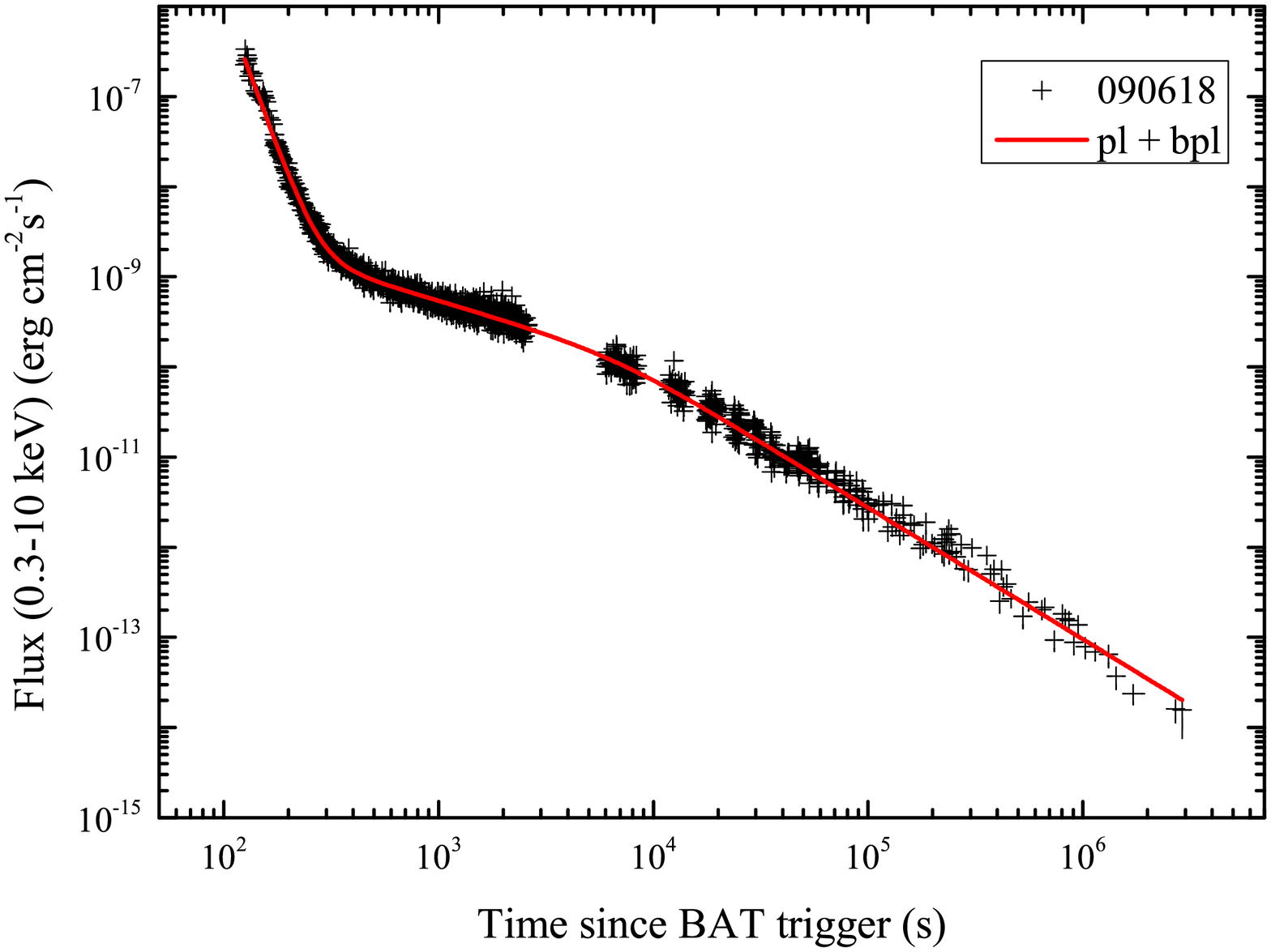} &
\includegraphics[angle=0,scale=0.18,trim=10 0 10 0,clip]{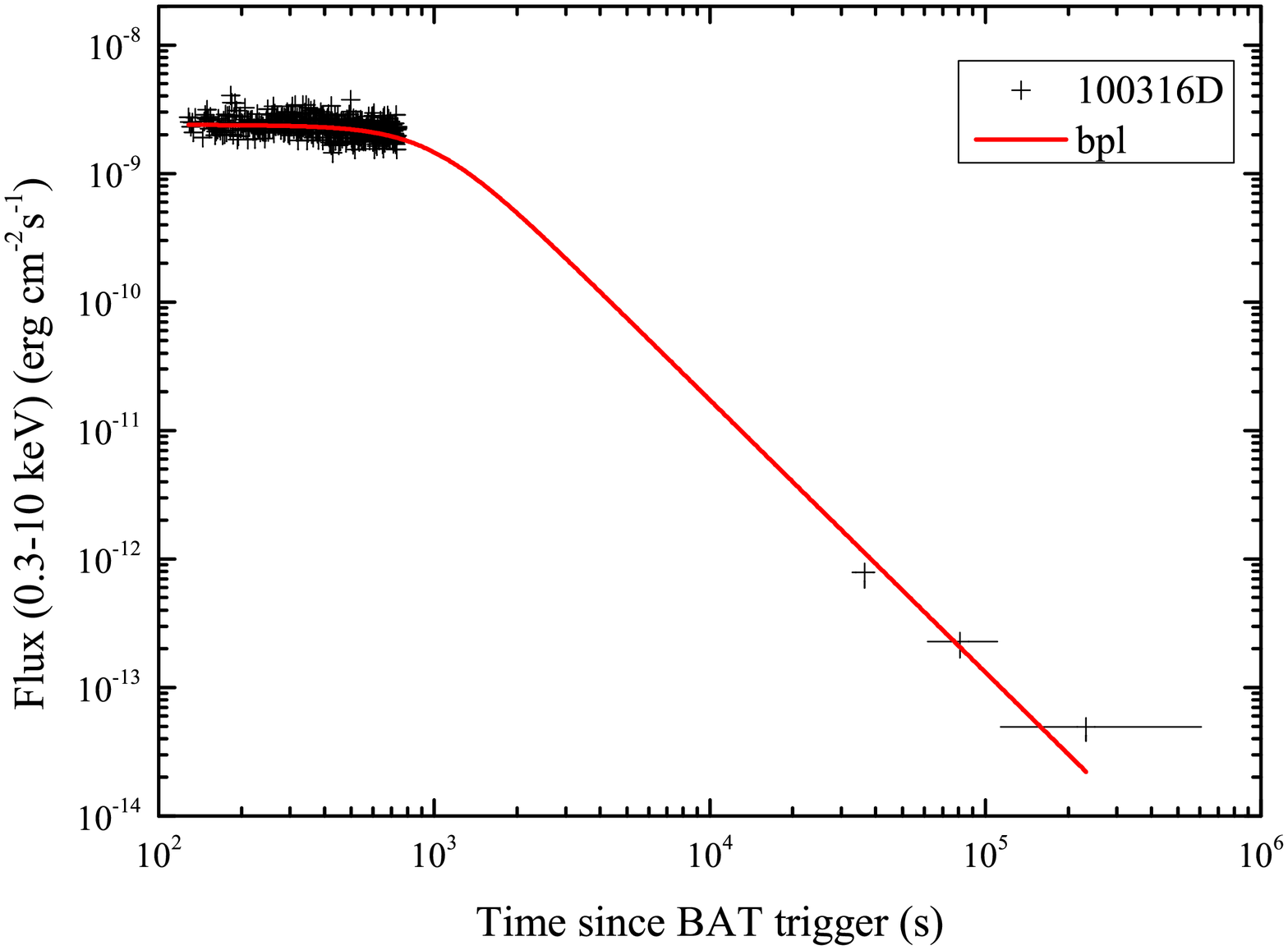} \\
\includegraphics[angle=0,scale=0.18,trim=10 0 10 0,clip]{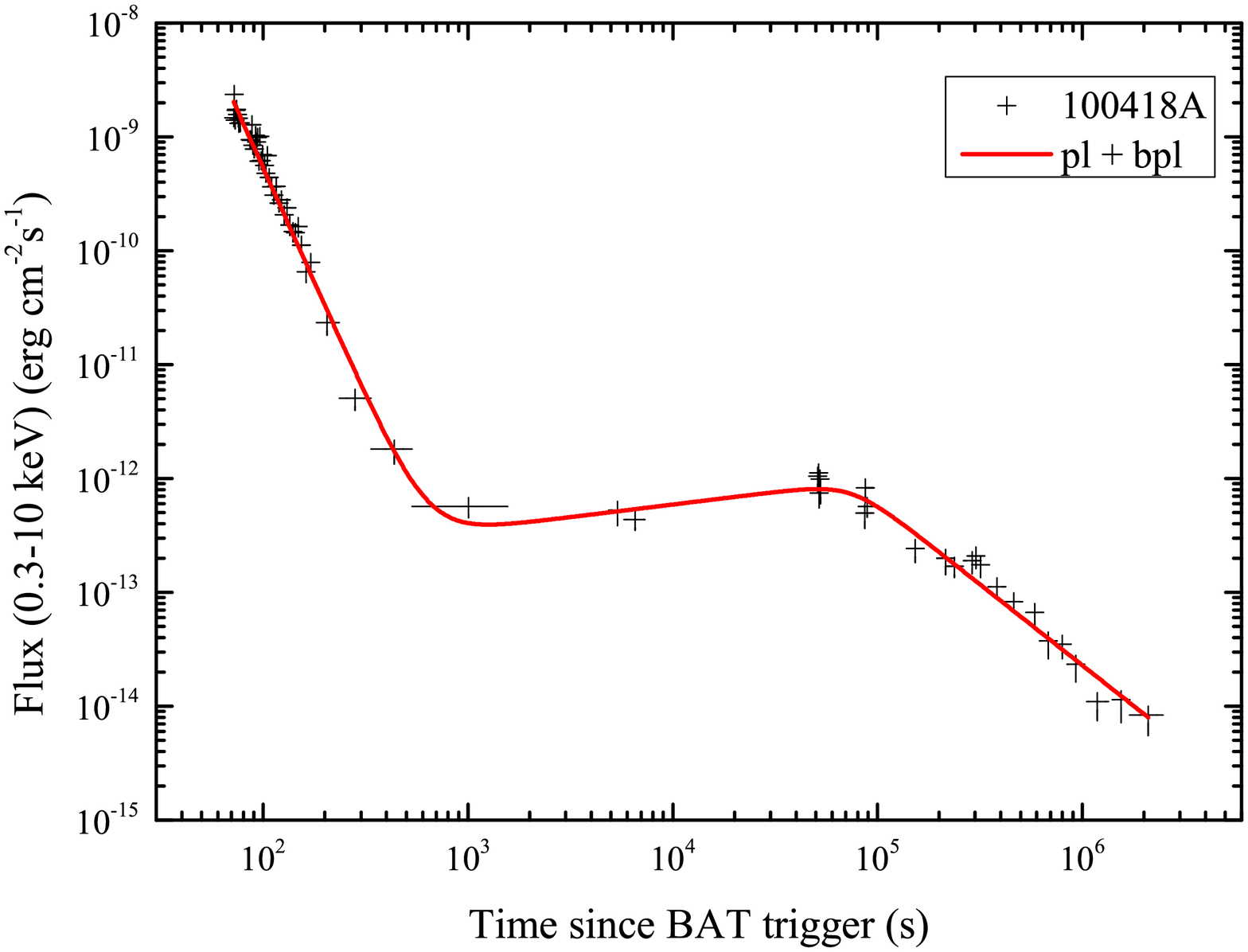}&
\includegraphics[angle=0,scale=0.18,trim=10 0 10 0,clip]{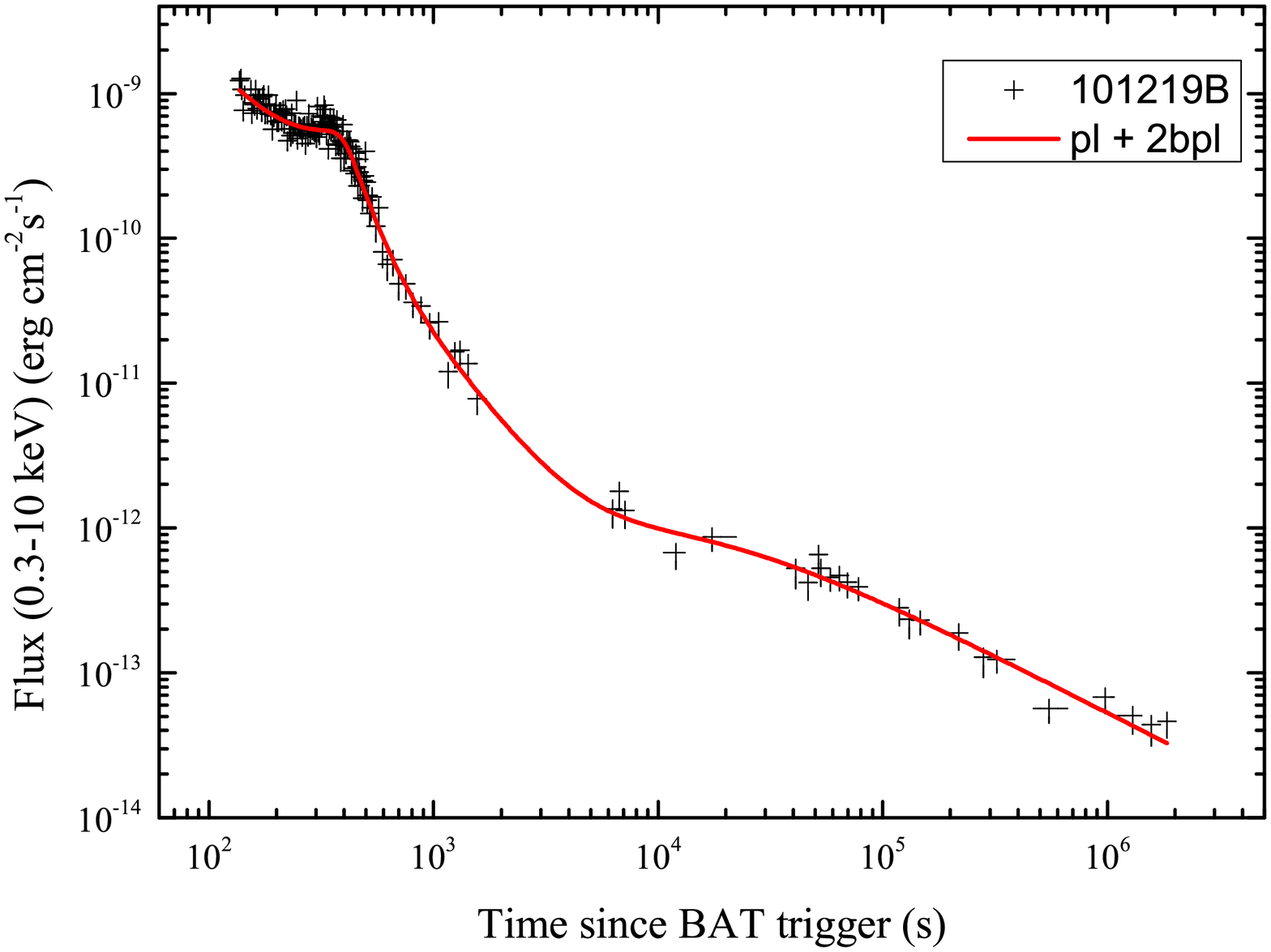} &
\includegraphics[angle=0,scale=0.18,trim=10 0 10 0,clip]{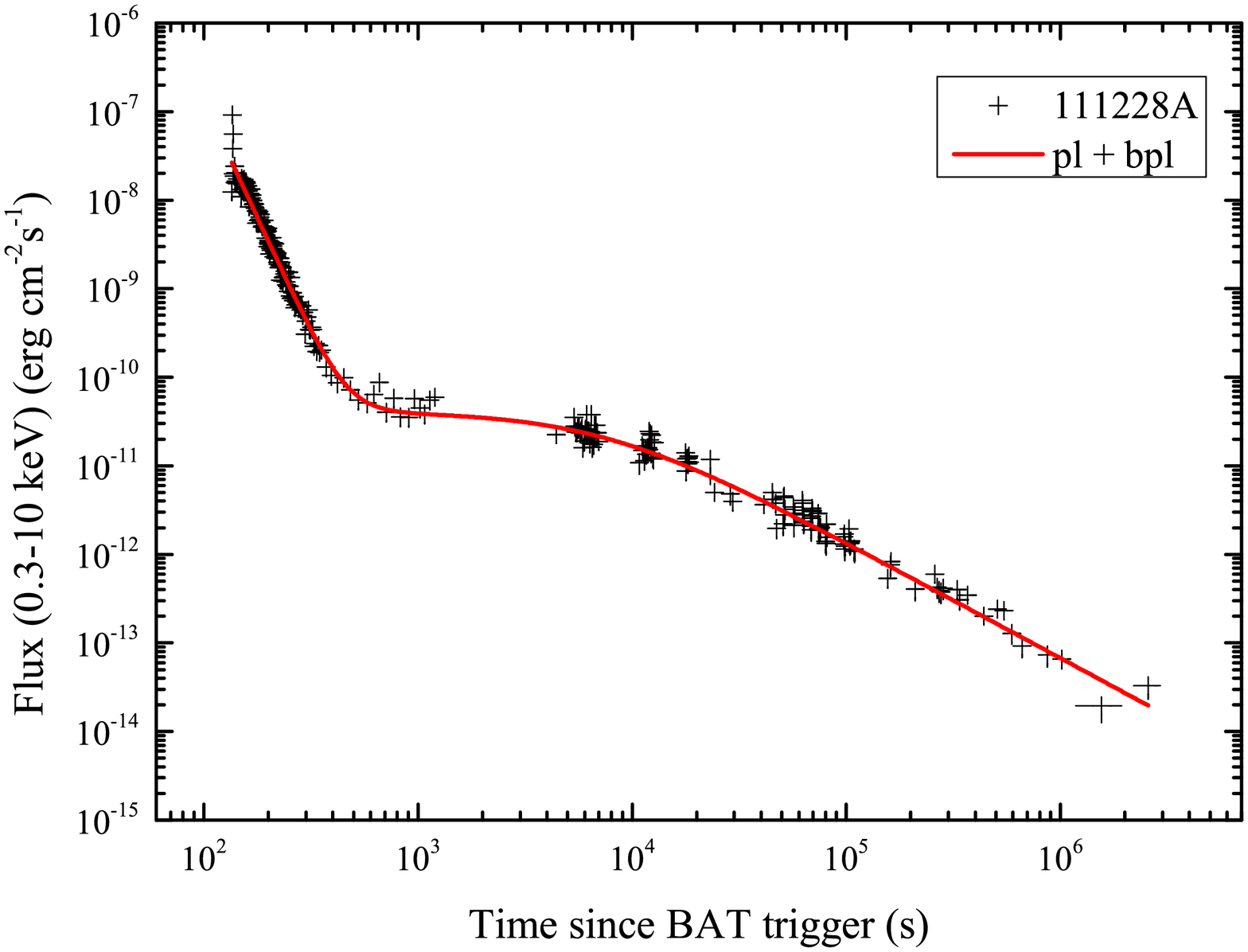}\\
\includegraphics[angle=0,scale=0.18,trim=10 0 10 0,clip]{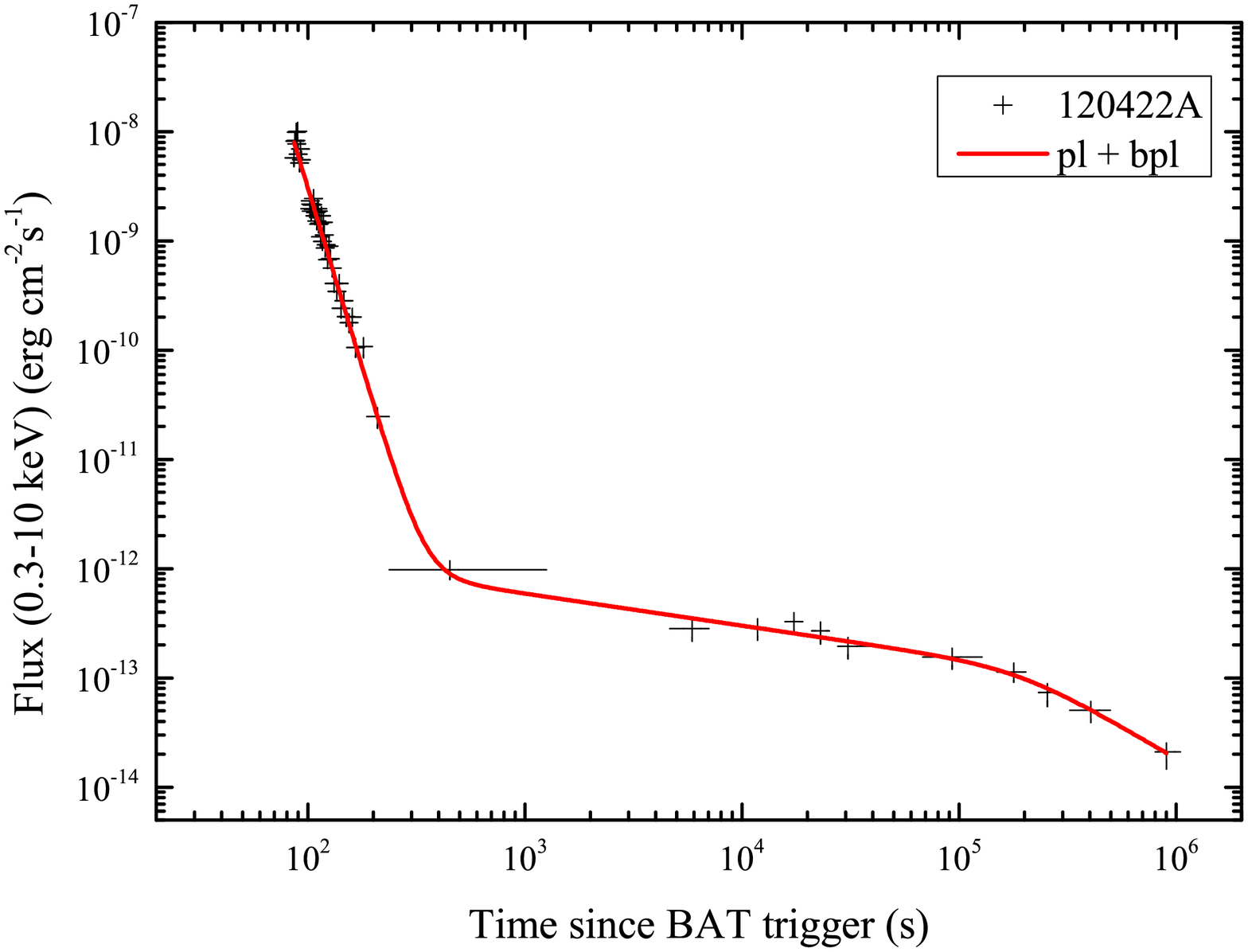}&
\includegraphics[angle=0,scale=0.18,trim=10 0 10 0,clip]{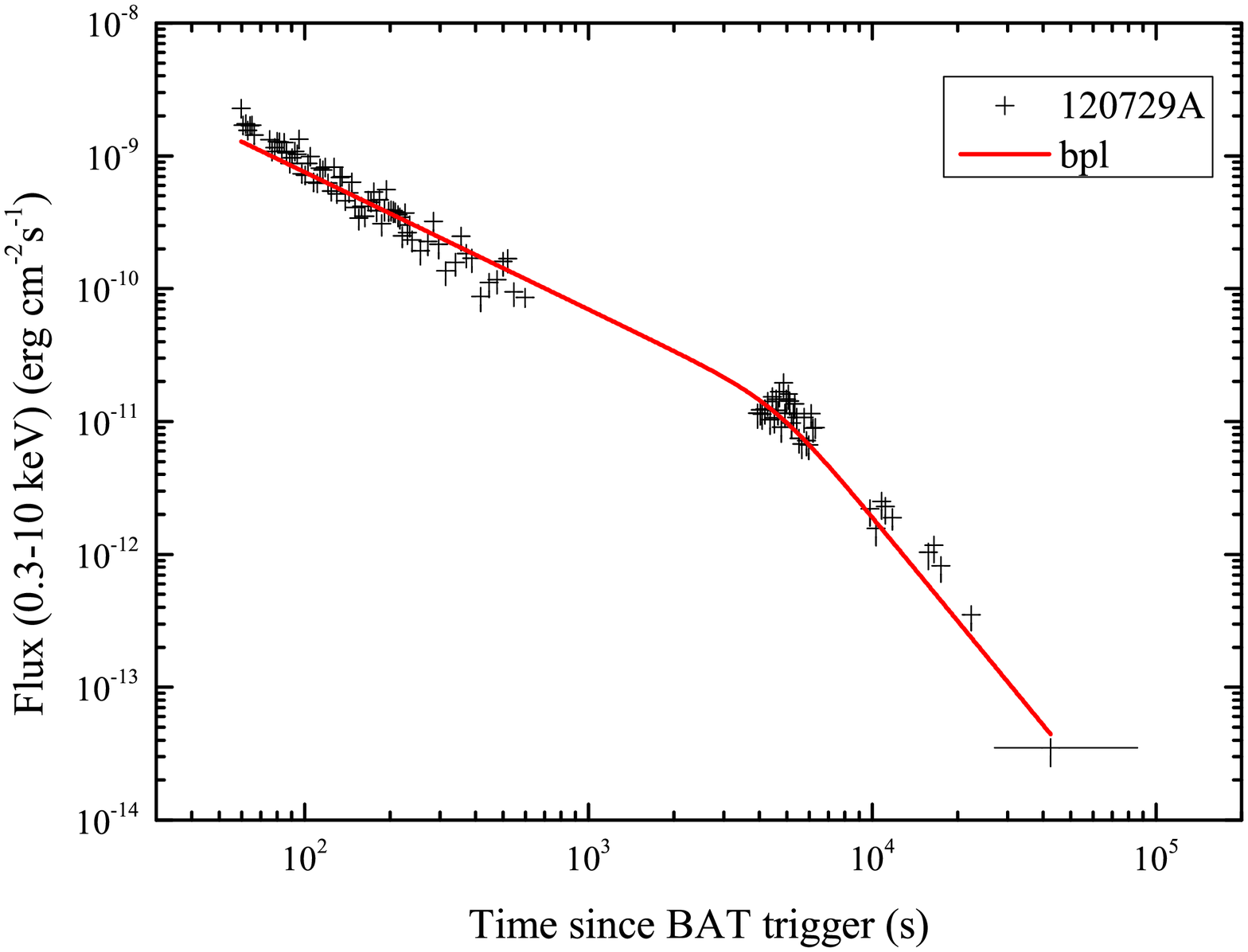}&
\includegraphics[angle=0,scale=0.18,trim=10 0 10 0,clip]{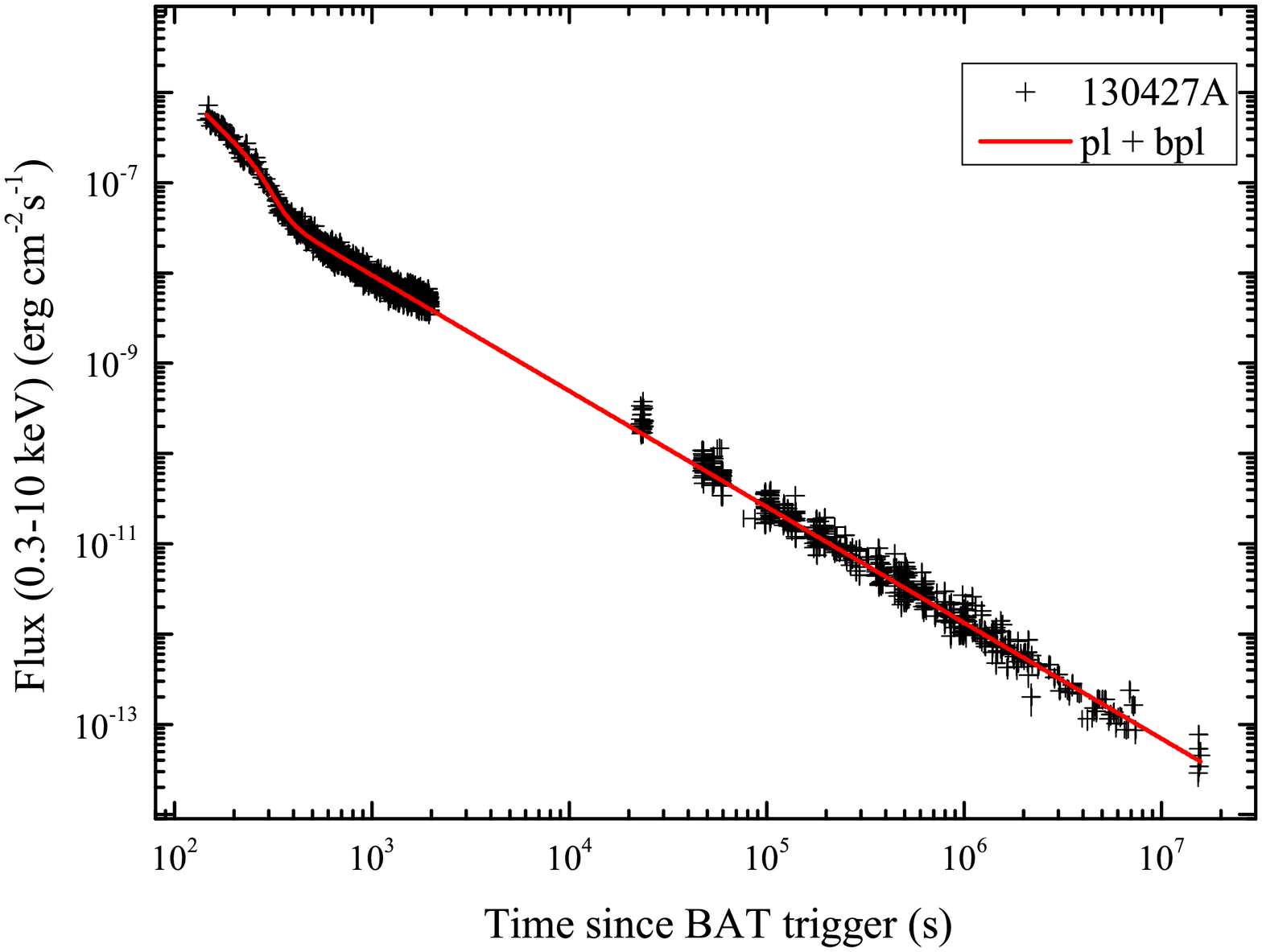}\\
\includegraphics[angle=0,scale=0.18,trim=10 0 10 0,clip]{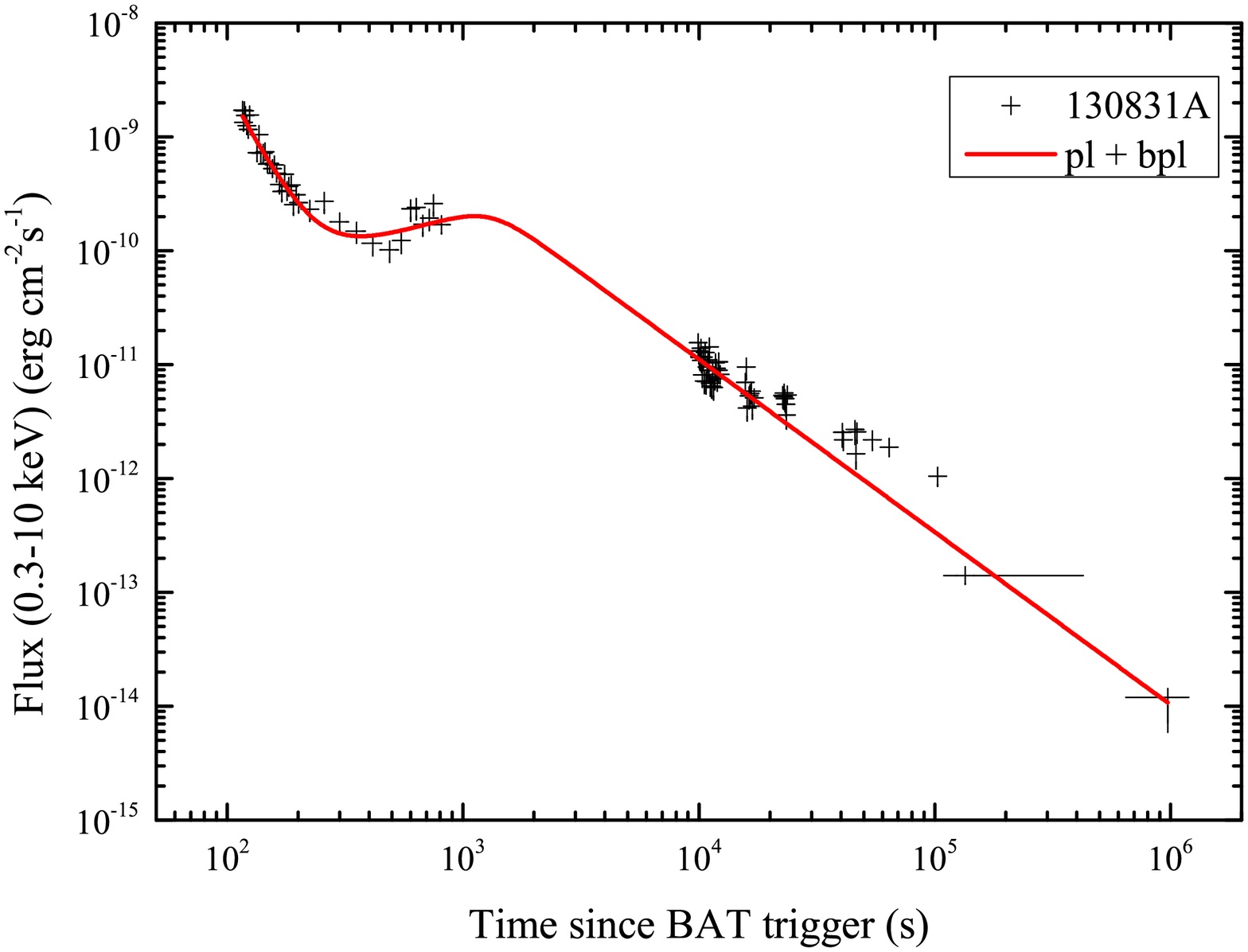}&
\includegraphics[angle=0,scale=0.18,trim=10 0 10 0,clip]{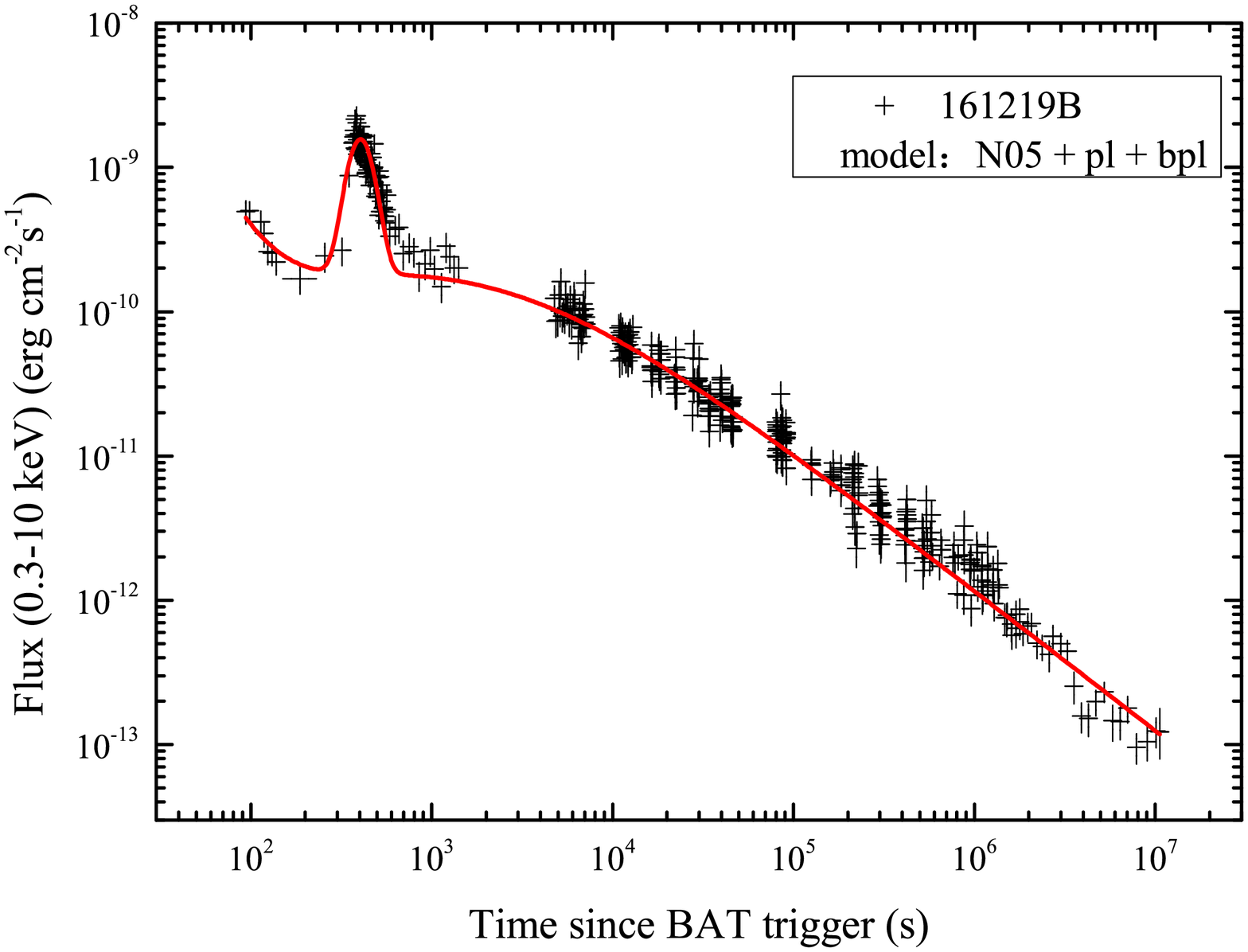}&
\includegraphics[angle=0,scale=0.18,trim=10 0 10 0,clip]{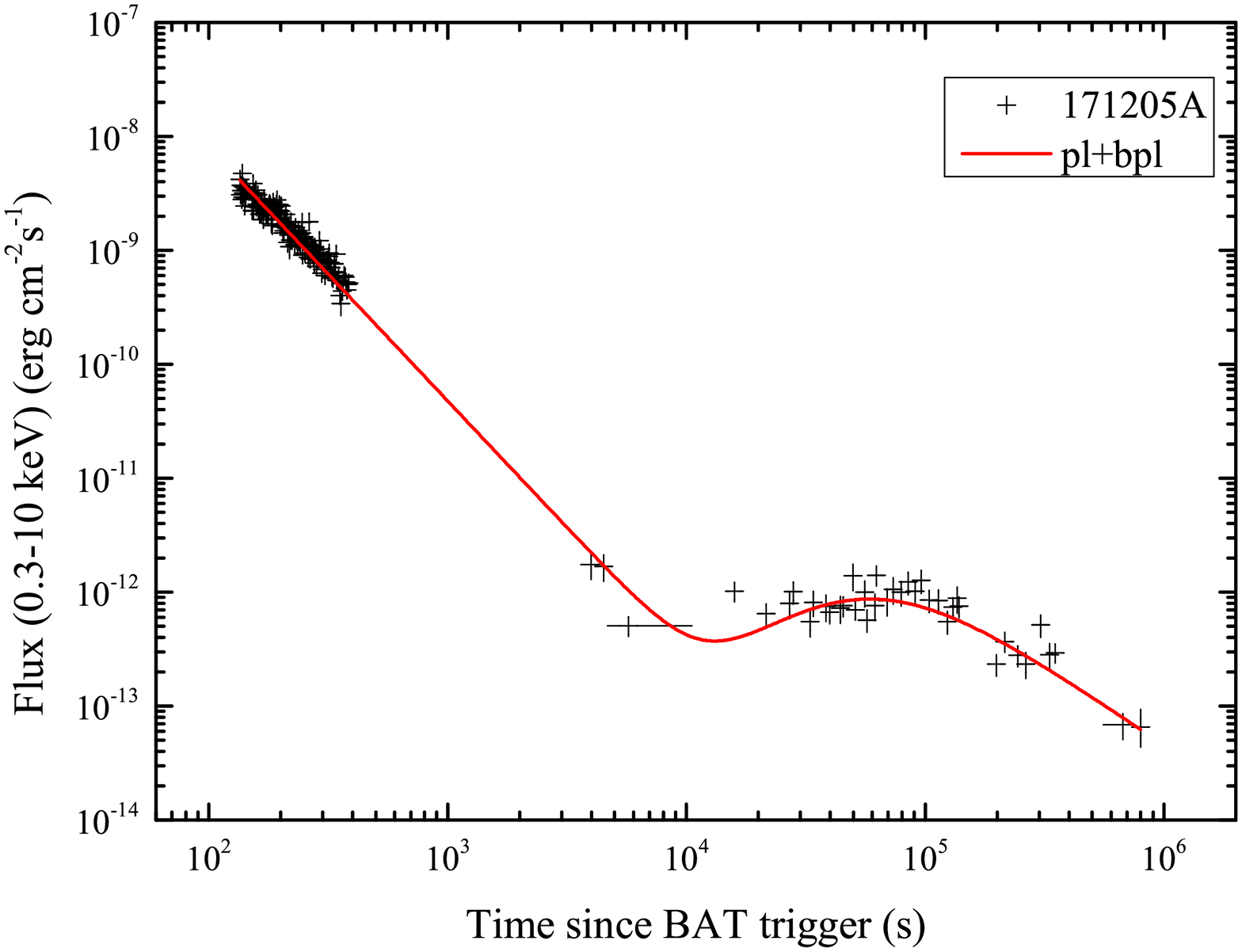}\\
\end{tabular}
\caption[] {The X-ray light curves of 18 SN-GRBs with early
\textit{Swift}/XRT follow-up observations in our Sample~\textrm{I}.
The best fitting is shown by the red curves.
The smooth broken power-law and single power-law are abbreviated to
``bpl" and ``pl", respectively.
 ``N05"  \citep{Norris2005} model means function~(\ref{N052}) in
 the fitting of the bright flares from GRBs 060904B and 161219B.
}
\label{F1}
\end{figure*}

The XRT light curves of all the 18 SN-GRBs in our Sample~\textrm{I}
\footnote{\url{http://www.swift.ac.uk/xrtcurves/} \citep{Evans2007,Evans2009}
and \url{http://www.astro.caltech.edu/grbox/grbox.php}}
are presented in Figure~\ref{F1}.
In this figure, a smooth broken power-law or a single power-law
\citep{Beuermann1999} was used to fit the light curve, and the 
related fitting parameters are reported in Table~\ref{T2}.
We examined the 18 GRBs and searched for bright X-ray flares
satisfying the condition ``$F_{\p}> 3F$" \citep[e.g.,][]{Yi2016,Mu2016a},
where $F_{\p}$ and $F$ are the peak flux and
the underlying continuum flux at the peak time of the flare, respectively.
We found that only two sources, GRBs 060904B and 161219B,
have bright X-ray flares, as shown by the mark ``B" (bright) in
the fourth column of Table~\ref{T1}.
Then, the occurrence rate of bright X-ray flares in our Sample~\textrm{I}
(SN-GRB) is only 11.1\%, which seems to be lower than
that in the general population.

\begin{table*}
\scriptsize
\caption{Fitting parameters of the 18 SN-GRBs in our Sample~\textrm{I}.
From left to right: $\rm{GRB}$, the power-law function parameters
($F_{0}$ and power-law index $\alpha$);
the smooth broken power-law function parameters ($F_{1}$,
the indices before and after the break time: $\alpha_{1}$ and $\alpha_{2}$,
the break time of the flare $t_{\rm{b}}$,
the sharpness around peak flux in the flare light curve $w$).
The light curve of GRB 060218 is fitted by the sum
of two smooth broken power-laws, and therefore the parameters
are shown in two lines ($060218^{a}$ and $060218^{b}$).
The light curve of GRB 101219B is fitted by the sum
of two smooth broken power-laws and one simple power-law, and therefore
the parameters are shown in two lines (101219B$^{a}$ and 101219B$^{b}$)}.
\begin{tabular}{lcccccccc}
\toprule
$\rm{GRB}$ & $F_{0}$ & $\alpha$ & $F_{1}$ &  $\alpha_{1}$ &  $\alpha_{2}$ & $t_{\rm b}$ & $w$ & $\rm{red}-\chi^{2}$ \\
& $\rm{(erg~cm^{-2}~s^{-1})}$ & & $\rm{(erg~cm^{-2}~s^{-1})}$ & & & ({\rm ks})& &  \\
\midrule
$050416\rm A^{\rm XRF}$	&	19.5	$\pm$	201	 &	5.78	$\pm$	2.43	&(	1.52	$\pm$	0.41	)$\times$ $10^{-11}$ &	0.37	$\pm$	0.11	&	0.9	$\pm$	0.02	&	1.55 	$\pm$	0.59 	&	3	&	0.84	\\
$060218^{\rm XRF,a}$	&	--	&			--			&(	4.73	$\pm$	0.05	)$\times$ $10^{-9}$ &	-0.54	$\pm$	0.01	&	5.47	$\pm$	0.04	&	2.53 	$\pm$	0.02 	&	0.5	&	-- 	\\
$060218^{\rm XRF,b}$	&	--	&			--			&(	1.52	$\pm$	0.27	)$\times$ $10^{-12}$ &	-12.7	$\pm$	10.08	&	1.16	$\pm$	0.06	&	11.55 	$\pm$	1.12 	&	0.5	&	1.32	\\
060729	&(	1.10 	$\pm$	0.39 	)$\times$ $10^{4}$ &	5.39	$\pm$	0.07	&(	1.42	$\pm$	0.07	)$\times$ $10^{-11}$ &	0.09	$\pm$	0.03	&	1.38	$\pm$	0.02	&	62.21 	$\pm$	3.01 	&	3	&	2.17	\\
060904B	&	--	&			--			&(	7.79	$\pm$	5.32	)$\times$ $10^{-12}$ &	0.6			&	1.42	$\pm$	0.15	&	5.58 	$\pm$	3.73 	&	3	&	8.71	\\
070419A	&	0.27 	$\pm$	0.13 	&	3.64	$\pm$	0.11	&(	2.40 	$\pm$	2.12	)$\times$ $10^{-13}$ &	-7.58	$\pm$	3.45	&	0.55	$\pm$	0.15	&	0.93 	$\pm$	0.99 	&	3	&	0.78	\\
080319B	&(	1.84 	$\pm$	0.05 	)$\times$ $10^{-4}$ &	1.6	$\pm$	0.11	&	--			&	--			&	--			&	--			&	--	&	2.09	\\
081007	&	7.46 	$\pm$	1.15 	 &	4.83	$\pm$	0.34	&(	1.67	$\pm$	0.63	)$\times$ $10^{-12}$ &	1.26	$\pm$	0.07	&	0.66	$\pm$	0.03	&	40.37 	$\pm$	16.61 	&	3	&	0.97	\\
090618	&(	1.27 	$\pm$	0.41 	)$\times$ $10^{7}$ &	6.52	$\pm$	0.07	&(	1.43	$\pm$	0.09	)$\times$ $10^{-10}$ &	0.7	$\pm$	0.01	&	1.46	$\pm$	0.01	&	6.65 	$\pm$	0.39 	&	3	&	1.06	\\
$100316\rm D^{\rm XRF}$	&	--	&			--			&(	2.36	$\pm$	0.04	)$\times$ $10^{-9}$ &	0.01	$\pm$	0.03	&	2.12			&	0.98 	$\pm$	0.09 	&	1.5	&	1.05	\\
100418A	&(	5.66 	$\pm$	2.51 	)$\times$ $10^{-2}$ &	4.01	$\pm$	0.12	&(	9.21	$\pm$	1.61	)$\times$ $10^{-13}$ &	-0.22	$\pm$	0.1	&	1.42	$\pm$	0.09	&	74.35 	$\pm$	14.85 	&	3	&	1.54	\\
101219B$^{a}$	&(	9.59 	$\pm$	5.07 	)$\times$ $10^{-6}$ &	1.90	$\pm$	0.09	&(	4.57	$\pm$	0.28	)$\times$ $10^{-10}$ &	-0.74 $\pm$ 0.33 			&	5.43	$\pm$	0.56	&	0.39	$\pm$	0.01 	&	3	&	--	\\
101219B$^{b}$	&-- &	--	&(	2.79	$\pm$	1.08	)$\times$ $10^{-12}$ &	-1.38 $\pm$ 1.66 			&	0.79	$\pm$	0.14	&	6.85 	$\pm$	3.66 	&	0.5	&	1.05	\\
111228A	&(	4.43 	$\pm$	1.68 	)$\times$ $10^{3}$ &	5.26	$\pm$	0.08	&(	4.16	$\pm$	0.77	)$\times$ $10^{-11}$ &	-0.01	$\pm$	0.1	&	1.31	$\pm$	0.05	&	7.36 	$\pm$	1.80 	&	1	&	1.88	\\
120422A	&(	4.34 	$\pm$	3.37 	)$\times$ $10^{4}$ &	6.58	$\pm$	0.18	&(	1.26	$\pm$	0.56	)$\times$ $10^{-13}$ &	0.29	$\pm$	0.08	&	1.18	$\pm$	0.42	&	193.4 	$\pm$	133.3 	&	3	&	1.57	\\
120729A	&	--	&			--			&(	1.44	$\pm$	0.08	)$\times$ $10^{-11}$ &	2.6	$\pm$	0.39	&	1.03		&	4.60 	$\pm$	0.75 	&	3	&	2.01	\\
130427A	&(	6.71 	$\pm$	0.12 	)$\times$ $10^{-5}$ &	1.28	$\pm$	0.01	&(	7.64	$\pm$	2.13	)$\times$ $10^{-8}$ &	2.39 $\pm$ 0.25 			&	10.1	$\pm$	1.3	&	0.31 	$\pm$	0.02 	&	1	&	1.34	\\
130831A	&	0.11 	$\pm$	0.30 	 &	3.82	$\pm$	0.5	&(	2.45	$\pm$	0.59	)$\times$ $10^{-10}$ &	-0.59	$\pm$	0.4	&	1.52	$\pm$	0.05	&	1.30 	$\pm$	0.23	&	3	&	1.94	\\
161219B	&(	1.75 	$\pm$	2.03 	)$\times$ $10^{-5}$ &	2.42	$\pm$	0.01	&(	6.37	$\pm$	3.07	)$\times$ $10^{-10}$ &	-0.5	$\pm$	0.73	&	0.97	$\pm$	0.1	&	1.54 	$\pm$	1.66 	&	0.5	&	8.74	\\
171205A	&(	2.24 	$\pm$	0.39 	)$\times$ $10^{-4}$ &	2.22 	$\pm$	0.03 	&(	3.39	$\pm$	0.31 	)$\times$ $10^{-12}$ &	-1.9	$\pm$	0.81	&	1.43	$\pm$	0.3	&	49.88 	$\pm$	19.65 	&	0.5	&	1.69	\\
\bottomrule
\end{tabular}
\label{T2}
\end{table*}

Since the detection of SNe is limited by the distance, secure SNe
identification becomes difficult because the SNe appears fainter
at a higher redshift \citep[see, e.g.,][]{Woosley2006}.
The SN-GRBs in Table~\ref{T1} exhibit redshift $z<1$ whereas the median
redshift of Swift long GRBs is above $2$ \citep{Jakobsson2006,Salvaterra2012}.
Thus, for a comparison, we studied another sample, Sample~\textrm{II},
which consists of all the $z<1$ LGRBs between January 2005 and
December 2017 that satisfy the aforementioned three criteria,
except for the 18 sources in Sample~\textrm{I}.
In other words, Sample~\textrm{II} is composed of the $z<1$ LGRBs
with rapid and adequate XRT follow-up observations, and without
optically observed SNe association.
We should stress that, here the words ``without observed SNe association"
do not mean ``SN-less". In our opinion, many SN-GRBs may still exist in
our Sample II. However, due to observational constraints, accompanying
SNe was not observed for those GRBs. The potential influence of such an
issue on our results is discussed in the last section.
Furthermore, \citet{Hjorth2012} showed that SN-less GRBs,
such as GRBs 060505 \citep{Fynbo2006}, 060614 \citep{Fynbo2006,Della2006,
Gal2006}, and possibly 051109B \citep{Perley2006} and XRF 040701
\citep{Soderberg2005}, have been observed to deep limits and no accompanying
SN is found. Among these sources,
only GRBs 051109B and 060614 have XRT rapid follow-up observations,
and have adequate observational data in the early time.
Since a macronova is likely to be associated with GRB 060614 \citep{Yang2015},
which indicates a merger of two compact objects,
060614 is not included in Sample~\textrm{II}.
Our Sample~\textrm{II} comprises 45 GRBs (44 LGRBs and one XRF),
see Table~\ref{T3} for details.
Among all of these 45 GRBs, we found that 16 GRBs
have bright X-ray flares, which are denoted as ``B" (bright)
in the third column of Table~\ref{T3}.

On the other hand, some dim X-ray fluctuations may
be regarded as weak X-ray flares.
Here we define the dim X-ray fluctuation as the
condition ``$1.5F<F_{\p}< 3F$" is satisfied.
Thus, only one SN-GRB has dim X-ray fluctuation, as denoted by
the character ``D" (dim) in the fourth column of Table~\ref{T1}.
Furthermore, in Sample~\textrm{II}, nine GRBs have dim X-ray
fluctuations, as reported by ``D" (dim) in the third column of Table~\ref{T3}.
The number of flares, $N_{\rm flare}$, is also given in the
last column of Table~\ref{T3}.

\begin{table*}
\caption{Sample~\textrm{II} of 45 $z<1$ GRBs.
From left to right: 44 LGRBs and one XRF (091018); redshift $z$;
bright X-ray flares and dim X-ray fluctuations denoted by ``B" (bright)
and ``D" (dim), respectively; number of flares.}
\centering
\begin{tabular}{lccc}
\toprule
 $\rm{GRB}$ &$z$ & Flare & $N_{\rm flare}$ \\
\midrule
050219A	&	0.2115	&	--	&	--	\\
050826	&	0.297	&	--	&	--	\\
051016B	&	0.9364	&	--	&	--	\\
051109B	&	0.08	&	--	&	--	\\
060202	&	0.785	&	D	&	1	\\
060512	&	0.4428	&	B	&	1	\\
060814	&	0.84	&	D	&	1	\\
060912A	&	0.937	&	--	&	--	\\
061021	&	0.3463	&	--	&	--	\\
061110A	&	0.758	&	--	&	--	\\
070318	&	0.84	&	B	&	1	\\
070508	&	0.82	&	--	&	--	\\
070521	&	0.553	&	--	&	--	\\
071112C	&	0.8227	&	B	&	1	\\
080430	&	0.767	&	--	&	--	\\
080916A	&	0.689	&	--	&	--	\\
081109	&	0.9787	&	--	&	--	\\
090424	&	0.544	&	--	&	--	\\
090814A	&	0.696	&	--	&	--	\\
$091018^{\rm XRF}$	&	0.971	&	--	&	--	\\
100508A	&	0.5201	&	D	&	1	\\
100621A	&	0.542	&	--	&	--	\\
100816A	&	0.804	&	B	&	1	\\
110715A	&	0.82	&	D	&	1	\\
111225A	&	0.297	&	D	&	1	\\
120722A	&	0.9586	&	B	&	1	\\
120907A	&	0.97	&	D	&	1	\\
130925A	&	0.347	&	B	&	5	\\
131103A	&	0.599	&	B	&	3	\\
140506A	&	0.889	&	B	&	3	\\
140512A	&	0.725	&	B	&	1	\\
140710A	&	0.558	&	B	&	1	\\
141004A	&	0.57	&	--	&	--	\\
150323A	&	0.593	&	D	&	1	\\
150727A	&	0.313	&	--	&	--	\\
150821A	&	0.755	&	B	&	1	\\
151027A	&	0.81	&	B	&	1	\\
160117B	&	0.86	&	B	&	2	\\
160131A	&	0.97	&	--	&	--	\\
160314A	&	0.726	&	B	&	1	\\
160425A	&	0.555	&	B	&	2	\\
160804A	&	0.736	&	D	&	1	\\
161129A	&	0.645	&	--	&	--	\\
170519A	&	0.818	&	B	&	1	\\
170607A	&	0.557	&	D	&	1	\\

\bottomrule
\end{tabular}
\label{T3}
\end{table*}

\section{Fitting procedure}\label{sec:fit}

\begin{figure*}
\begin{tabular}{cc}
\includegraphics[angle=0,scale=0.3,trim=30 0 30 0,clip]{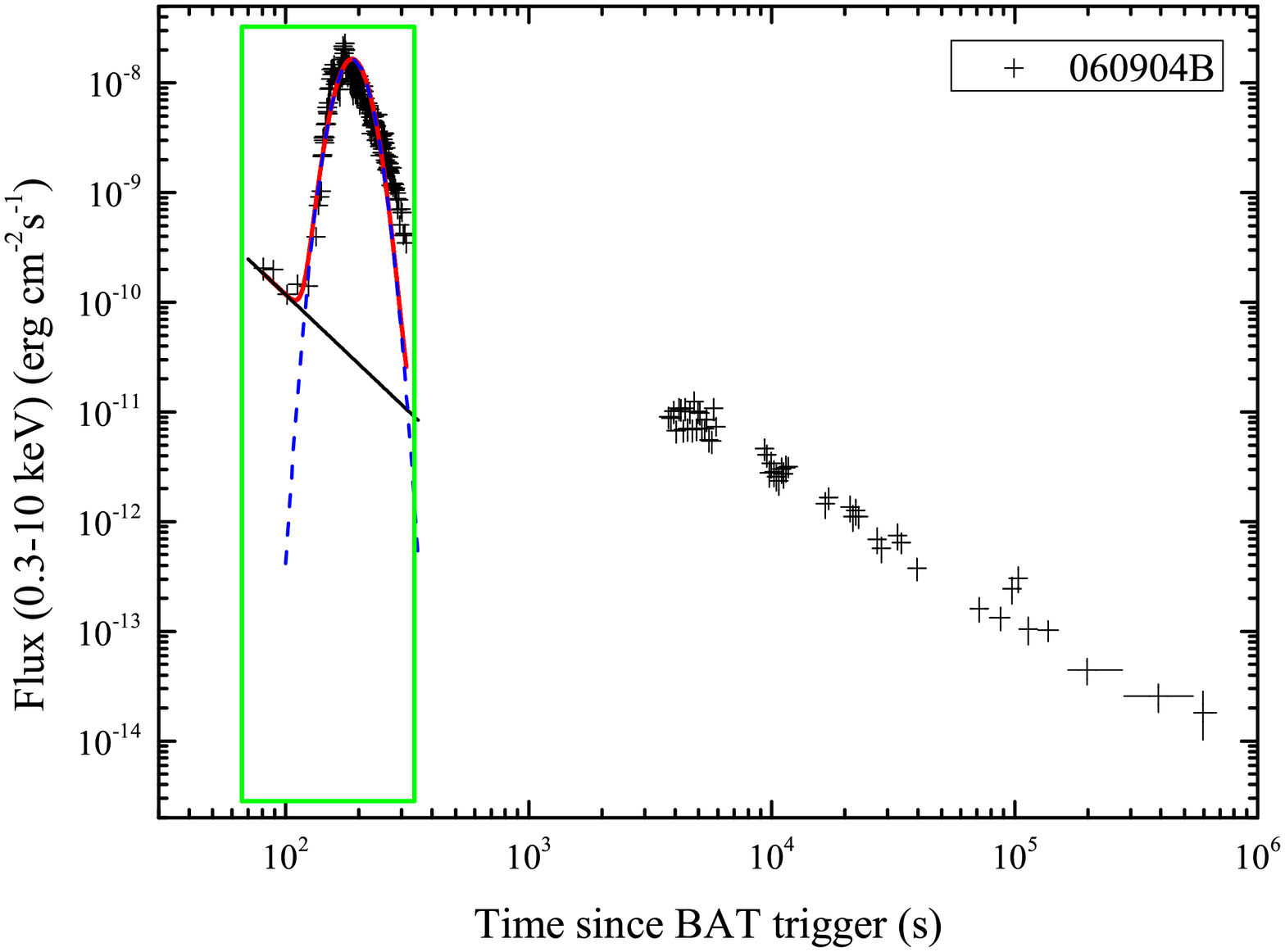}&
\includegraphics[angle=0,scale=0.3,trim=30 0 30 0,clip]{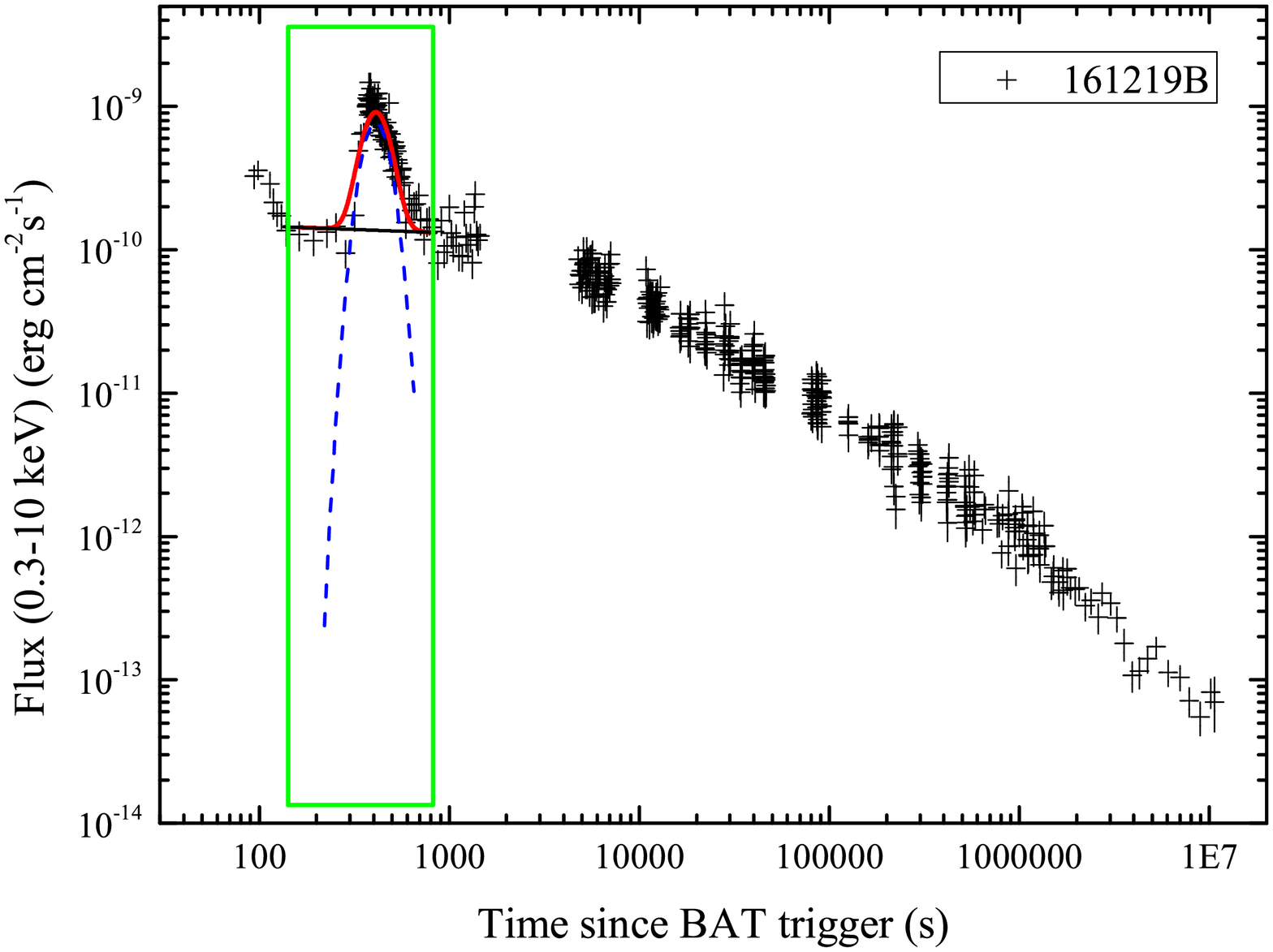}\\
\end{tabular}
\caption{Best fitting for the $0.3-10~\keV$ X-ray afterglow
light curve of GRBs 060904B and 161219B (red solid line).
Blue dashed line: best fitting for the bright X-ray flare;
black solid line: best fitting for the continuous X-ray emission
of the underlying continuum.
The green box marks the interval of the flare in our fitting.}
\label{F2}
\end{figure*}

The X-ray flare properties were investigated by fitting the
0.3 $-$ 10 \keV (\Swift/XRT) light curve with an empirical
function proposed by \citet{Norris2005}, for $t\geqslant t_{\rm s}$,
\begin{equation}
F_{t}=A\lambda e^{-\frac{\tau_{1}}{t-t_{\rm{s}}}
- \frac{t-t_{\rm{s}}}{\tau_{2}}} \ ,
\label{N051}
\end{equation}
where $\mu=(\tau_{1}/\tau_{2})^{1/2}$ and $\lambda=e^{2\mu}$.
The peak time of the flare is $t_{\p}=\tau_{\p}+t_{\s}
= (\tau_{1}\tau_{2})^{1/2}+t_{\s}$.
The time of flare onset $t_{\s} = 0$ is adopted in Equation~(\ref{N051}).
Then, the above equation is simplified as
\begin{equation}
F_{t}=A e^{2(\tau_{1}/\tau_{2})^{1/2}
- \frac{\tau_{1}}{t}-\frac{t}{\tau_{2}}} \ .
\label{N052}
\end{equation}
Thus, the peak flux of flare is $A=F_{\rm{max}}=F_{\tau_{\p}}$,
and $\tau_{\p}=(\tau_{1}\tau_{2})^{1/2}$.
The flare width is measured between the two $1/e$ intensity points,
\begin{equation}
\omega=\Delta t_{1/e}=\tau_{2}(1+4\mu)^{1/2}.
\label{width}
\end{equation}
The fitting results of the X-ray flares are reported in Table~\ref{T4}.
In Sample~\textrm{I}, each of GRBs 060904B and 161219B has a single
bright flare. In Sample~\textrm{II}, 26 bright flares exist in 16 GRBs,
among which five GRBs have multiple flares and 11 GRBs have a single flare,
as shown in the last column of Table~\ref{T3}.
In this work, our concerns focus on whether or not, the
central engine exhibits reactivity behaviour.
As investigated by \citet{Bernardini2011},
a sizable fraction of late-time flares
(i.e. those with peak time $t_{\p} \ga 10^{3}~\s$) are
compatible with afterglow
variability. On the contrary, the early flares are more
likely to be related to central engine reactivity.
If a GRB has multiple X-ray flares, it is reasonable
to judge whether the central engine becomes
reactive after the prompt emission
by studying the physical origin of the first flare.
Thus, for a GRB with multiple flares,
only the first flare was fitted. The fitting procedure of
GRBs 060904B and 161219B is shown in Figure~\ref{F2}.

In addition, to estimate the relative variability
flux $\Delta F/F$, where $\Delta F$ and $F$ are the increase
of the flux and the underlying continuum flux at the peak
time of the flare, respectively,
then, the underlying continuum was fitted by using
a simple power-law (black solid line, Fig.~\ref{F2})
\citep[e.g.,][]{Bernardini2011,Margutti2011}.
The fitting results of the underlying continuum are
reported in Table~\ref{T4}.
The quantities $t^{\rm{rest}}_{\p}$ and $\omega^{\rm{rest}}$ are
also shown in Table~\ref{T4}, where $t^{\rm{rest}}_{\p} = t_{\p}/(1+z)$
and $\omega^{\rm{rest}} = \omega/(1+z)$ are the peak time and width of
flares in the rest frame, respectively.

\begin{table*}
\scriptsize
\caption{Physical parameters from the flares in Sample~\textrm{I}
and Sample~\textrm{II}, and the underlying continuum fitting.
From left to right: $\rm{GRB}$, amplitude $\rm{A}$ and the shape parameters
$\tau_{1}$, $\tau_{2}$ of the best-fitting of bright X-ray flare from
\citet{Norris2005};
$t^{\rm{rest}}_{\p}$ and $\omega^{\rm{rest}}$ are the peak
time and width of flares in the rest frame, respectively.
$F_{0}$ and $\alpha$ are the parameters of the underlying continuum.}
\centering
\begin{tabular}{lcccccccc}
\toprule
 $\rm{GRB}$ & $\rm{A}$ & $\tau_{1}$ & $\tau_{2}$ & $t^{\rm{rest}}_{\p}$ & $\omega^{\rm{rest}}$ & $F_{0}$ & $\alpha$ & $\rm{red}-\chi^{2}$  \\
 &  $\rm{(erg~cm^{-2}~s^{-1})}$ & ($10^{3}$~{\s}) & ({\s}) &  ({\s}) &  ({\s}) & $\rm{(erg~cm^{-2}~s^{-1})}$  &  & \\
\midrule
 \multicolumn{9}{c}{Sample~\textrm{I}}\\
\midrule
060904B	&(	1.65 	$\pm$	0.05 	)$\times$ $10^{-8}$ &	4.86 	$\pm$	0.13 	&	7.23 	$\pm$	0.19 	&		110.1 	$\pm$	3.6 	&		43.47 	$\pm$	0.29 		&(	1.86 	$\pm$	0.62 	)$\times$ $10^{-6}$ &	2.10 	 	&	8.17 	\\
161219B	&(	7.76 	$\pm$	0.34 	)$\times$ $10^{-10}$ &	8.27 	$\pm$	0.89 	&	20.38 	$\pm$	2.07 	&		357.7 	$\pm$	30.4 	&		160.4 	$\pm$	1.3 		&(	1.85 	$\pm$	0.18 	)$\times$ $10^{-10}$ &	0.05 	$\pm$	0.01 	&	2.27 	\\
\midrule
\multicolumn{9}{c}{Sample~\textrm{II}}\\
\midrule																																	
060512	&(	2.47 	$\pm$	0.38 	)$\times$ $10^{-10}$ &	6.39 	$\pm$	1.85 	&	7.72 	$\pm$	2.21 	&		154.0 	$\pm$	45.2 	&		57.68 	$\pm$	4.06 		&(	1.25 	$\pm$	0.14 	)$\times$ $10^{-6}$ &	1.80 	$\pm$	0.23 	&	1.56 	\\
070318	&(	5.90 	$\pm$	0.28 	)$\times$ $10^{-10}$ &	3.64 	$\pm$	0.36 	&	24.19 	$\pm$	2.25 	&		161.2 	$\pm$	20.2 	&		93.01 	$\pm$	0.45 		&(	1.12 	$\pm$	0.18 	)$\times$ $10^{-7}$ &	1.07 	$\pm$	0.03 	&	1.50 	\\
071112C	&(	2.76 	$\pm$	0.18 	)$\times$ $10^{-10}$ &	22.78 	$\pm$	4.87 	&	19.34 	$\pm$	3.97 	&		364.2 	$\pm$	98.4 	&	124.8 	$\pm$	2.8 		&(	5.75 	$\pm$	0.46 	)$\times$ $10^{-7}$ &	1.35 	$\pm$	0.01 	&	1.30 	\\
100816A	&(	3.30 	$\pm$	0.48 	)$\times$ $10^{-10}$ &	3.31 	$\pm$	1.03 	&	6.58 	$\pm$	1.98 	&		81.8 	$\pm$	31.9 	&		34.75 	$\pm$	2.69 		&(	1.17 	$\pm$	0.98 	)$\times$ $10^{-7}$ &	1.29 	$\pm$	0.17 	&	1.01 	\\
120722A	&(	5.64 	$\pm$	0.67 	)$\times$ $10^{-11}$ &	2.86 	$\pm$	1.60 	&	33.00 	$\pm$	15.39 	&		156.9 	$\pm$	111.9 	&		104.2 	$\pm$	1.7 		&(	2.34 	$\pm$	16 	)$\times$ $10^{8}$ &	8.42 	$\pm$	1.51 	&	0.65 	\\
130925A	&(	4.14 	$\pm$	0.05 	)$\times$ $10^{-9}$ &	33.86 	$\pm$	0.90 	&	24.38 	$\pm$	0.63 	&		674.4 	$\pm$	16.8 	&	221.7 	$\pm$	0.5 		&(	5.35 	$\pm$	1.31 	)$\times$ $10^{-3}$ &	2.41 	$\pm$	0.05 	&	1.88 	\\
131103A	&(	1.54 	$\pm$	0.12 	)$\times$ $10^{-9}$ &	2.35 	$\pm$	0.38 	&	2.83 	$\pm$	0.44 	&		51.0 	$\pm$	9.2 	&	19.09 	$\pm$	2.02 		&(	1.20 	$\pm$	2.69 	)$\times$ $10^{-8}$ &	0.99 	$\pm$	0.43 	&	1.59 	\\
140506A	&(	3.80 	$\pm$	0.12 	)$\times$ $10^{-8}$ &	3.04 	$\pm$	0.08 	&	5.19 	$\pm$	0.14 	&		66.5 	$\pm$	2.4 	&	27.17 	$\pm$	0.25 		&(	1.24 	$\pm$	0.15 	)$\times$ $10^{-7}$ &	0.97 	$\pm$	0.01 	&	2.18 	\\
140512A	&(	1.50 	$\pm$	0.05 	)$\times$ $10^{-8}$ &	3.10 	$\pm$	0.19 	&	5.14 	$\pm$	0.28 	&		73.2 	$\pm$	5.2 	&	29.69 	$\pm$	0.59 		&(	7.30 	$\pm$	6.79 	)$\times$ $10^{-9}$ &	0.30 	$\pm$	0.18 	&	0.65 	\\
140710A	&(	1.22 	$\pm$	0.15 	)$\times$ $10^{-10}$ &	7.62 	$\pm$	1.88 	&	21.04 	$\pm$	4.92 	&		256.9 	$\pm$	68.2 	&	118.6 	$\pm$	2.1 		&(	6.63 	$\pm$	1.26 	)$\times$ $10^{-7}$ &	1.95 	$\pm$	0.39 	&	1.96 	\\
150821A	&(	1.56 	$\pm$	0.10 	)$\times$ $10^{-10}$ &	135.4 	$\pm$	41.7 	&	19.03 	$\pm$	5.73 	&		914.4 	$\pm$	345.8 	&	199.4 	$\pm$	10.4 		&(	1.04 	$\pm$	0.41 	)$\times$ $10^{-6}$ &	1.34 	$\pm$	0.04 	&	0.83 	\\
151027A	&(	4.50 	$\pm$	0.13 	)$\times$ $10^{-8}$ &	6.21 	$\pm$	0.20 	&	2.53 	$\pm$	0.08 	&		69.2 	$\pm$	2.9 	&	19.71 	$\pm$	0.63 		&(	1.36 	$\pm$	0.04 	)$\times$ $10^{-7}$ &	0.70 	$\pm$	0.01 	&	1.66 	\\
160117B	&(	3.24 	$\pm$	0.19 	)$\times$ $10^{-9}$ &	2.53 	$\pm$	0.40 	&	3.09 	$\pm$	0.46 	&		47.5 	$\pm$	9.6 	&	17.84 	$\pm$	1.67 		&(	8.43 	$\pm$	23 	)$\times$ $10^{4}$ &	7.48 	$\pm$	0.74 	&	3.41 	\\
160314A	&(	2.75 	$\pm$	0.50 	)$\times$ $10^{-11}$ &	1.25 	$\pm$	0.57 	&	83.55 	$\pm$	25.01 	&		187.5 	$\pm$	87.6 	&	196.6 	$\pm$	0.6 		&(	3.27 	$\pm$	1.16 	)$\times$ $10^{-5}$ &	2.78 	$\pm$	0.07 	&	0.98 	\\
160425A	&(	3.38 	$\pm$	0.15 	)$\times$ $10^{-8}$ &	17.60 	$\pm$	0.50 	&	5.36 	$\pm$	0.15 	&		197.5 	$\pm$	6.1 	&	52.29 	$\pm$	0.73 		&(	5.54 	$\pm$	3.75 	)$\times$ $10^{-5}$ &	2.07 	$\pm$	0.11 	&	7.97 	\\
170519A	&(	1.14 	$\pm$	0.15 	)$\times$ $10^{-8}$ &	8.16 	$\pm$	2.04 	&	5.70 	$\pm$	1.26 	&		118.7 	$\pm$	35.9 	&	38.72 	$\pm$	3.46 		&(	1.33 	$\pm$	20 	)$\times$ $10^{-5}$ &	1.73 	$\pm$	2.67 	&	6.80 	\\
\bottomrule
\end{tabular}
\label{T4}
\end{table*}

\section{Occurrence rates and physical origin of bright X-ray flares}\label{sec:CE}

This work focuses on the occurrence rates of bright X-ray flares
in the SN-GRB sample (Sample~\textrm{I}) and the general $z<1$
GRBs without observed SNe association (Sample~\textrm{II}).
As shown in Section~\ref{sec:sample}, for Sample~\textrm{I},
among the 18 SN-GRBs (15 LGRBs and three XRFs), only two SN-GRBs
have bright X-ray flares, and the occurrence rate is 11.1\%.
For a comparison, for Sample~\textrm{II},
among the 45 GRBs (44 LGRBs and one XRF),
16 sources present bright X-ray flares, and the occurrence rate is 35.6\%.
Thus, the occurrence rate of X-ray flares in the SN-GRB systems is
lower than that in Sample~\textrm{II}.
In addition, such a discrepancy between these two samples
can be examined by the Fisher's exact test
\footnote{\url{http://www.langsrud.com/stat/fisher.htm}},
which shows the one-tailed $P = 0.0466 \ (< 0.05)$.
On the other hand, if the dim X-ray fluctuation is included as
the weak flare, then 16.7\% (3/18) of Sample~\textrm{I} and
55.6\% (25/45) of Sample~\textrm{II} have X-ray flares,
again showing the discrepancy between these two samples.
Moreover, the Fisher's exact test shows the one-tailed
$P = 0.0048 \ (\ll  0.05) $.
Thus, the discrepancy may indicate that the SN-GRB systems have a lower
occurrence rate of X-ray flares than the general $z<1$ GRB systems without
observed SNe association. In other words, a lower occurrence rate of
X-ray flares may exist in the SN-GRB sample than in the general
$z<1$ GRB population.

\begin{figure}
\centering
\hspace*{-10mm}
\includegraphics[width=\columnwidth]{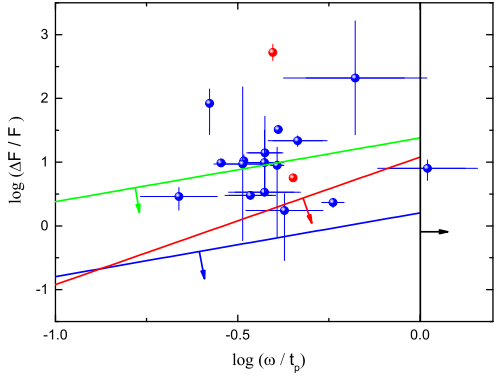}
\caption{
Relationship between the relative variability flux $\Delta F/F$
and the relative variability timescale $\omega /t_{\p}$ for
the X-ray flares in our Sample~\textrm{I} and Sample~\textrm{II}.
The two bright flares from Sample~\textrm{I}, which are denoted
by the red circles. The 16 bright flares from Sample~\textrm{II}
are denoted by the blue circles.
The four theoretical solid lines are identical with those in
Figure~6 of \citet{Bernardini2011}, i.e., density fluctuations
on axis (blue line) and off-axis (red line), off-axis multiple regions
density fluctuations (green line), patchy shell model (black line),
see \citet{Ioka2005} for details.}
\label{F3}
\end{figure}

It should be noted that the physics of X-ray flares may be based on
the internal origin or the external one. \citet{Ioka2005} showed
that simple kinematic arguments can give limits on the timescale
$\omega$ and amplitude $\Delta F$ of variabilities in GRB afterglows.
They proposed that four kinds of afterglow variability
are kinematically forbidden under some standard assumptions, and derived
the limits for dips (bumps) that deviate below (above) the baseline with a
timescale and amplitude (see their Figure~1 for details).
These limits are helpful to identify whether or not, the physical origin is
afterglow variability or the late-time activity of the central engine.
Similar to \citet{Ioka2005}, \citet{Bernardini2011},
and \citet{Mu2016a}, we plot a figure based on the relative variability
flux $\Delta F/F$ and the relative variability timescale
$\omega /t_{\rm p}$ to judge the physical origin of the X-ray flares.
As shown in Figure~\ref{F3}, most flares are located in the upper
left region, which indicates that they are likely to be of internal origin.

\begin{figure}
\centering
\hspace*{-10mm}
\includegraphics[width=\columnwidth]{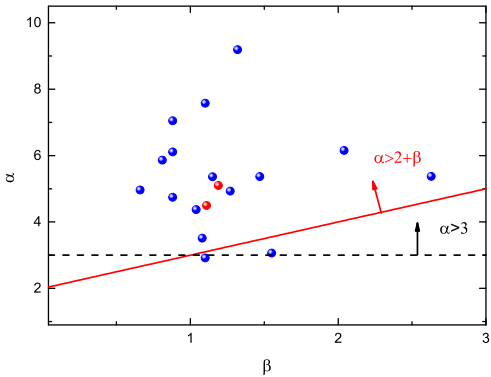}
\caption{
A comparison of the X-ray flares in our two samples with the
internal origin $``\alpha > 2 + \beta"$ (the region above the red
solid line) and $``\alpha >3"$ (the region above the black dashed line).
The flares from Sample~\textrm{I} and Sample~\textrm{II} are denoted by
the red and blue circles, respectively.
The meaning of red and blue circles are the same as in Figure~\ref{F3}.
}
\label{F4}
\end{figure}

In another way, by setting the zero time at the GRB trigger time,
the flares formed in the external shock
process may have a maximum decay slope of $``\alpha = 2 + \beta"$,
where $\beta$ is the spectral index.
Consequently, a decay with a slope steeper than $2+\beta$,
(i.e., $``\alpha > 2 + \beta"$), may indicate the internal origin
\citep[e.g.,][]{Kumar2000,Liang2006}.
Such a simple criterion can also be simplified as an even simpler
one, $``\alpha >3"$, since $\beta$ is usually around 1.
The light curve index $\alpha$ in the decay phase
(from $t_{\p}$ to $t_{\p} + \tau_{\rm{dec}}$)
can be roughly estimated as
\begin{equation}\label{lcdec}
\alpha=\frac{\log (e)}{\log [(t_{\p} + \tau_{\rm dec})/t_{\p}]} \ ,
\end{equation}
where $\tau_{\rm{dec}}=\tau_{2}[(1+4\mu)^{1/2}+1]/2$ is the decay time
of the flare. The temporal decay index is listed in Table~\ref{T5}.
The spectral analyses for the steep decay segments are
performed by using a power-law spectral model
\footnote{\url{http://www.swift.ac.uk/xrt spectra/addspec.php/}},
The spectral analyses results, i.e., the values of the spectral index
in the decay phase $\beta$, are reported in Table~\ref{T5}.
It is seen from Figure~\ref{F4} that, most flares are located above
the red solid line and the black dashed line, which implies that
most flares are likely to be of internal origin.
Such a result is in good agreement with the data shown in Figure~\ref{F3}.
We therefore argue that most X-ray flares
studied in this work are related to the reactivity of the central engine
\citep{Romano2006,Bernardini2011,Wu2013,Yi2015}.

\begin{table*}
\centering
\caption{Parameters for the physical origin of Sample~\textrm{I} and
Sample~\textrm{II}.
From left to right: $\rm{GRB}$; $\omega/t_{\p}$ is the relative
time-scale, where $\omega$ is the width evaluated between $1/e$
intensity points, and $t_{\p}$ is the peak time of the flare;
$\Delta F/F$ is the relative flux at $t_{\p}$, and
$F$ is calculated from the best fitting of the underlying continuum;
$\alpha$ is the temporal decay index;
$\beta$ is the average spectral index in the decay phase.
}

\begin{tabular}{lcccc}
\toprule
$\rm{GRB}$ & $\omega/t_{\p}$ & $\Delta F/F$ & $\alpha$ &  $\beta$  \\
\midrule
 \multicolumn{5}{c}{Sample~\textrm{I}}\\
\midrule
 060904B & 0.39 $\pm$	0.01 & 527 $\pm$ 176 & 5.10 & 1.19\\
 161219B & 0.45 $\pm$	0.03 & 5.67 $\pm$ 0.56 &	4.50 &1.11\\
\midrule
 \multicolumn{5}{c}{Sample~\textrm{II}}\\
\midrule
060512	&	0.37 	$\pm$	0.08 	&	3.38 	$\pm$	0.38 	&	5.37 	&	2.63	\\
070318	&	0.58 	$\pm$	0.04 	&	2.33 	$\pm$	0.38 	&	3.51 	&	1.08	\\
071112C	&	0.34 	$\pm$	0.05 	&	3.01 	$\pm$	0.24 	&	5.87 	&	0.81	\\
100816A	&	0.42 	$\pm$	0.09 	&	1.74 	$\pm$	1.45 	&	4.74 	&	0.88	\\
120722A	&	0.66 	$\pm$	0.24 	&	210 	$\pm$	143 	&	3.07 	&	1.55	\\
130925A	&	0.33 	$\pm$	0.01 	&	10.50 	$\pm$	2.57 	&	6.11 	&	0.88	\\
131103A	&	0.37 	$\pm$	0.04 	&	9.86 	$\pm$	22.13 	&	5.37 	&	1.47	\\
140506A	&	0.41 	$\pm$	0.01 	&	32.54 	$\pm$	3.91 	&	4.93 	&	1.27	\\
140512A	&	0.41 	$\pm$	0.02 	&	8.93 	$\pm$	8.25 	&	4.96 	&	0.66	\\
140710A	&	0.46 	$\pm$	0.08 	&	21.77 	$\pm$	4.14 	&	4.37 	&	1.04	\\
150821A	&	0.22 	$\pm$	0.05 	&	2.89 	$\pm$	1.12 	&	9.19 	&	1.32	\\
151027A	&	0.28 	$\pm$	0.01 	&	9.74 	$\pm$	0.31 	&	7.05 	&	0.88	\\
160117B	&	0.38 	$\pm$	0.04 	&	14.00 	$\pm$	6.08 	&	5.36 	&	1.15	\\
160314A	&	1.05 	$\pm$	0.28 	&	7.99 	$\pm$	2.84 	&	2.92 	&	1.10    \\
160425A	&	0.26 	$\pm$	0.01 	&	83.75 	$\pm$	56.66 	&	7.58 	&	1.10	\\
170519A	&	0.33 	$\pm$	0.05 	&	9.36 	$\pm$	8.81 	&	6.16 	&	2.04	\\
\bottomrule
\end{tabular}
\label{T5}
\end{table*}

\section{Discussion and Conclusions}\label{sec:Conclusions}

This work focuses on the different occurrence rates of bright X-ray
flares in the $z<1$ GRBs with (Sample~\textrm{I}) or
without (Sample~\textrm{II}) observed SNe association.
Our Sample~\textrm{I} consists of 18 SN-GRBs,
among which only two GRBs have bright X-ray flares.
Sample~\textrm{II} consists of 45 GRBs,
among which 16 GRBs have bright X-ray flares.
Our study has shown a lower occurrence rate of bright X-ray flares
in the SN-GRB sample (2/18, 11.1\%) than in
Sample~\textrm{II} (16/45, 35.6\%).
In addition, if the dim X-ray fluctuation is included as a dim flare,
then 16.7\% (3/18) of the SN-GRB systems and 55.6\% (25/45) of
Sample~\textrm{II} have flares, again showing the discrepancy between
these two samples.
Thus, the discrepancy may indicate that a lower occurrence rate of
X-ray flares may exist in the SN-GRB sample than in the general
$z<1$ GRB population.

It is known that there exists a strong selection effect of distance on
the luminosity and the total energy of GRBs. In the present work, however,
we focus on those close GRBs with $z<1$ , so the selection effect of
distance may not be essential.
To our knowledge, none of the known selection effects
seems to play a role that could account for the apparent deficit
of flares in the SN-GRB sample.

As mentioned in the second section, owing to observational constraints,
many SN-GRBs may still exist in our Sample II.
In addition, we should point out that our work is based on the assumption
that there may exist a group of bona fide SN-less LGRBs.
Otherwise, our arguments
as well as the division of Samples I and II will make less sense.
If Sample II does consist of two groups, i.e., SN-GRBs (Sample IIa)
and bona fide SN-less GRBs (Sample IIb),
we would argue that our main results, i.e., the discrepancy on the
occurrence of X-ray flares, can still work.
The arguments are as follows. We assume that
Samples I, IIa, and IIb have $N_1$, $N_2$, and $N_3$ sources, respectively.
If there indeed exists a discrepancy on the occurrence
rate of X-ray flares between the SN-GRBs and the bona fide SN-less LGRBs,
we assume that the occurrence rate is $f_1$ for the former and
$f_2$ for the latter, with $f_1 < f_2$.
Obviously, the real difference in the rates is $(f_2-f_1)$.
In such case, according to our analyses based on Samples I and II,
the apparent difference will be
$[(f_1*N_2 + f_2*N_3)/(N_2 + N_3)] - f_1 = [N_3/(N_2+N_3)]\times (f_2-f_1)$,
which is even lower than the real one, i.e., $(f_2-f_1)$.
In other words, if there exists an apparent discrepancy between
Samples I and II, such a discrepancy is likely to be even more significant
in the real case (between Sample IIb and Sample I plus IIa).
We therefore argue that even though it is not clear how many SN-GRBs
exist in our Sample II, the discrepancy suggested in this work may still have
potential significance.

In our opinion, the physical understanding of the lower rate of occurrence
in the SN-GRB systems may be the following. From the view of the energy source,
both the radiation of an SNe and the prompt gamma-ray emission
together with the X-ray flare, originate from the total energy of
collapse of a massive star, which can be roughly regarded as
a saturated energy budget.
To produce bright flares, one needs to have in-falling
materials near the equatorial direction, which is in the opposite
sense of the outgoing materials to power an SN. Thus, bright flares
may mean more in-falling materials and therefore less materials
are ejected to power the SNe.
If this is the case, then it is understandable why the SN-GRB
sample has a lower occurrence rate of X-ray flares than
Sample~\textrm{II} population.

We should point out that, the above argument on the saturated energy budget
is only qualitative. In this work, we did not calculate the energy of X-ray
flares even though the fitting parameters have been obtained.
The reasons for this are as follows.
First, a typical ratio of the isotropic energy of a flare to GRB is
around 10\% \citep[e.g.,][]{Chincarini2010,Yi2016}, which means that
the radiation energy of an X-ray flare is generally less than that of
the prompt gamma-ray emission.
However, the difference in the opening angle and in the
energy-release efficiency between the gamma-ray emission and
the corresponding flare remains uncertain.
It may be less sense to simply add up the two aforementioned parts
of the energy budget.
In addition, it is difficult to estimate the total energy released
through an individual SN, where the neutrino radiation is dominant.
Thus, it is beyond the scope of the present
work to conduct a detailed energy evaluation. That is why we just focused on
whether or not, the central engine has late-time reactivity, and just fitted
the first flare but did not investigate, in further detail, the amount of
released energy.

On the other hand, the millisecond magnetar may also work as the central
engine for GRBs
\citep[e.g.,][]{Usov1992,Duncan1992,Dai1998,Zhang2001,Metzger2015}.
It is known that the total amount of available energy of a magnetar
is around $2\times 10^{52}~\erg$
\citep[e.g.,][]{Thompson2004,Metzger2011},
and perhaps up to $10^{53}~\erg$ \citep{Metzger2015}.
In addition, it has been shown that the average kinetic energy of SN-GRBs
is around $2\times 10^{52}~\erg$ \citep{Mazzali2014,Cano2015},
which has implications for the total energy budget.
Thus, it is worth undertaking further investigation of the total energy
of SN-GRBs to study the type of central engines.
A recent interesting work related to this issue, \citet{Li2017},
investigated the total energy budget of X-ray plateaus and suggested that
a black hole is likely to be operating for most GRBs, and a magnetar
central engine is possible for 20\% of their analysed GRBs.

\section*{Acknowledgements}

We thank Bing Zhang and Xue-Feng Wu for beneficial discussion,
and thank the referee for helpful suggestions that improved the manuscript.
We acknowledge the use of the public data from the \textit{Swift}
data archive, and the UK \textit{Swift} Science Data Center.
This work was supported by the National Basic Research Program of China
(973 Program) under grants 2014CB845800,
and the National Natural Science Foundation of China under grants 11573023,
11473022, 11333004, 11673062, 11503011, 11773007, 11403005,
11473021, 11522323, 11525312, 11533003, and U1731239.
EWL acknowledges the special fundings for Guangxi distinguished
professors (Bagui Yingcai \& Bagui Xuezhe).
J.W. was supported by the Fundamental Research Funds for
the Central Universities under grant 20720160023.

\bsp
\label{lastpage}
\end{document}